\documentclass[aps,prl,superscriptaddress,twocolumn,showpacs]{revtex4-1}

\usepackage[utf8]{inputenc}
\usepackage[english]{babel}
\usepackage{graphicx} 
\usepackage{orcidlink}

\makeatletter\AtBeginDocument{\let\@elt\relax}\makeatother

\usepackage{amsmath}
\usepackage{amssymb}
\usepackage{xcolor}
\usepackage{color} 
\usepackage[abs]{overpic}
\usepackage{epstopdf}
\usepackage{soul}

\usepackage[normalem]{ulem}

\begin{document}

\title{Floquet time-crystals as quantum sensors of AC-fields}
\author{Fernando Iemini}
\affiliation{Instituto de F\'isica, Universidade Federal Fluminense, 24210-346 Niter\'oi, Brazil}
\affiliation{Institute for Physics, Johannes Gutenberg University Mainz, D-55099 Mainz, Germany}
\author{Rosario Fazio~\orcidlink{0000-0002-7793-179X}}
\affiliation{The Abdus Salam International Center for Theoretical Physics, Strada Costiera $11$, $34151$ Trieste, Italy}
\affiliation{Dipartimento di Fisica “E. Pancini", Università di Napoli “Federico II”, Monte S. Angelo, I-$80126$ Napoli, Italy}

\author{Anna Sanpera ~\orcidlink{0000-0002-8970-6127}}
\affiliation{Física Teòrica: Informació i Fenòmens Quàntics, Universitat Autònoma de Barcelona, $08193$ Bellaterra, Spain}
\affiliation{ICREA, Pg. Lluís Companys $23$, $08010$ Barcelona, Spain}

\date{\today}

\begin{abstract}


The long range spatial and temporal ordering displayed by discrete time crystals, can become advantageous properties when used for sensing extremely weak signals. Here, we investigate their performance as quantum sensors of weak AC-fields and demonstrate, using the quantum Fisher information measure, that they can overcome the shot noise limit while allowing long interrogation times. In such systems, collective interactions stabilize their dynamics against noise making them robust enough to protocol imperfections.

\end{abstract}


\maketitle
Time crystals are non-equilibrium phases of matter that break time-translation invariance and present long-range time ordering.
 Originally proposed by Wilczek in isolated systems~\cite{Wilczek2012}, it was soon realized that if energy is the only conserved quantity, time symmetry cannot be spontaneously broken in such cases~\cite{Watanabe2015}. Immediately after, several contributions demonstrated that time crystals may exist in out-of-equilibrium physics. Seminal theoretical investigations~\cite{Else2016,Khemani2016} put forward that periodically driven (Floquet) many-body interacting quantum systems can host time crystallinity, and its existence was soon after experimentally confirmed~\cite{Choi2017a,Zhang2017}. Since then, the field has become very active, see~\cite{Sacha_review,Else_review,Sacha-book,Zaletel2023}
and references therein for recent developments.

 A Floquet Time Crystal (FTC) corresponds to
a many-body interacting system driven by a periodic (Floquet)  Hamiltonian of period $T$, i.e., $H(t)=H(t+ T)$, that spontaneously breaks the \textit{discrete} time translation symmetry of the Hamiltonian. The FTC behaviour can be probed by measuring an order parameter $O(t)$  (e.g., global spin magnetization) that displays subharmonic response, i.e.,  $O(t)=O(t+nT)$ ($n\in\mathbb Z; n>1$). Crucially, such behaviour is stable against Hamiltonian perturbations, defining a real quantum phase of matter.

Beside the relevance of FTCs in understanding fundamental aspects of non-equilibrium quantum matter, time crystals may also offer advantages in applications. Investigations in this direction are, however, at an early stage ~\cite{Estarellas2020,Bomantara2018,Carollo2020,Montenegro2023,Cabot2023}. In this work we make a step forward in this direction by exploring their potential as quantum sensors.

Interacting many-body quantum systems may act as precise sensors of different physical quantities, from electric and magnetic fields, to frequency and temperature among others~\cite{Degen_review}. 
Moreover, the quantum correlations present in the system may result in a quantum-enhanced sensitivity. Such enhancement is reflected in the so-called quantum Fisher information that bounds the minimal uncertainty  with which an unknown parameter can be estimated, assuming one can optimize over all possible measurements.
Furthermore, one can exploit the criticality of phase transitions, 
 which enhances the correlation length of the system and its susceptibility to external fields~\cite{Zanardi2008,Invernizzi2008}, or alternatively use highly quantum entangled systems (e.g., GHZ-like states).
 In such cases, the Standard Quantum Limit (SQL) bounded by classical statistical correlations can be overcome, leading to the ultimate Heisenberg  limit precision~\cite{Giovannetti_review}.
However, in non-equilibrium dynamics, entanglement heats up the system as well, leading inevitably to noise and instabilities that will deteriorating the sensor performance.
 Thus, sensors exploiting the entanglement resources present in many-body interacting systems, while having an exponentially slow heating become robust to decoherence effects and are very promising for metrology.

 Here we use a spin system in a FTC phase as a quantum sensor of an AC-field, and analyze its performance as a function of the properties of the FTC.
 Our approach takes a different route from usual ones, which are mostly based on dynamical decoupling schemes~\cite{Khodjasteh2005,RevModPhys.88.041001,PhysRevX.10.031003}, integrability~\cite{Bayat2022} or prethermal stabilization of highly entangled states~\cite{Choi2017arxiv}.
 Before proceeding further let us briefly summarize our most relevant findings.  
The long-range spatial and time ordered dynamics present in the FTC, together with the presence of spin-spin correlations, allows for an optimal sensing performance for long times. Already for a moderate size of the sensor, its performance in time overcomes the standard quantum limit, even when estimating infinitesimal small AC-fields. The breaking of the time-translational invariance is crucial for boosting the performance of the sensor. A generic Floquet system, with either ergodic or non-ergodic dynamics, do not show similar advantages. Furthermore,  FTCs can also be robust in the presence of dissipation~\cite{Riera_Campeny2020} ensuring their capability as sensors.
\begin{figure}
\includegraphics[width = 0.78 \linewidth]{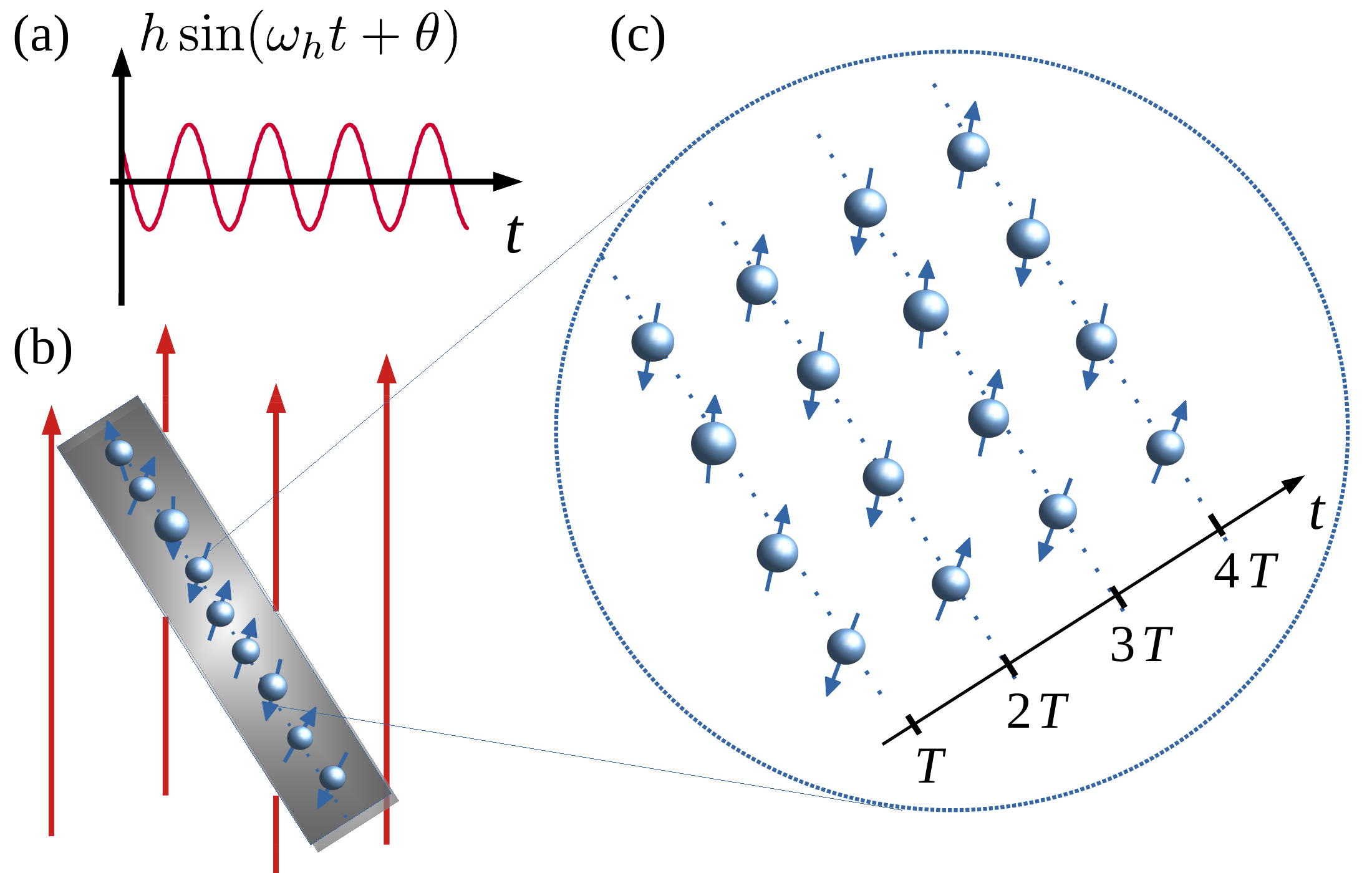}\\
\vspace{0.5cm}
\includegraphics[width = 0.78 \linewidth]{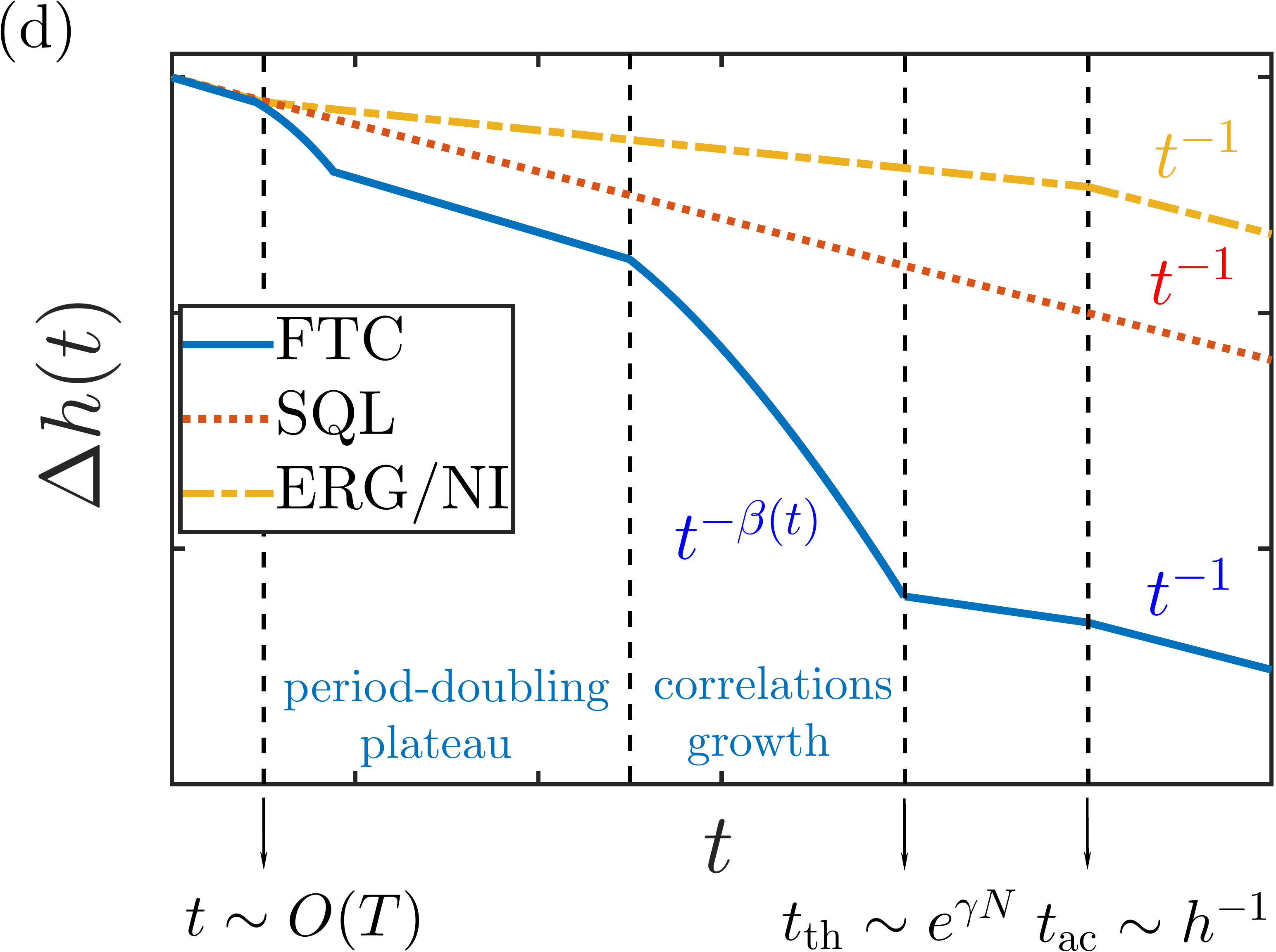}
\caption{ Sketch of the FTC quantum sensor: (a) external AC field of amplitude $h$ and frequency $\omega_h$.  (b-c) the sensor is composed by $N$ spins in a FTC, acting as a probe of the external AC-field. (d)  Variance for the optimal estimation of the amplitude $\Delta h (t)$ as a function of time for a generic ergodic or non-interacting phases (ERG/NI) and for a FTC phase. Due to its exponentially slow heating and many-body correlations, $\Delta h (t)$ in the FTC decreases faster than the SQL, $(\Delta h)_{\rm{SQL}}\sim 1/\sqrt{ (2t/\pi)^2 N }$ (\cite{SM}) until thermalization times $t_{th}$ are reached.
 }
\label{fig.sketch.introduction}
\end{figure}

\textbf{\textit{The model.}}
Our quantum sensor is composed of $N$  spins described by the time-dependent Hamiltonian

\begin{equation}
 \hat H_s = \sum_{i} \left[ J_{i} \hat \sigma_i^z \hat \sigma_{i+1}^z + \sum_{\alpha= x,z} b_{i}^\alpha \hat \sigma_i^\alpha -
\frac{\phi}{2} \sum_{n=-\infty}^\infty \delta(t-nT)   \hat{\sigma}_i^{x}\right]
\label{eq.model}
\end{equation}
 where $\hat \sigma_i^\alpha$ are Pauli operators  at site $i$ and component $\alpha$. The system is subject to periodic kicks at intervals $T$, inducing a global $\phi$-rotation. Between the kicks, the system evolves under disordered interactions and fields, with disorder strengths: $J_i\in [1/2,1/2 +J]$; $b_i^z\in[1/2,1/2+b^z]$; $b_i^x\in[0,b^x]$, chosen from a uniform distribution in the range of $J,b^z,b^x$. In the strong disordered regime and without kicking, the system evolves to a Many-Body Localized state (MBL), while the periodic kicking  at $\phi\simeq \pi$ stabilizes a FTC phase~\cite{Piotr_2023} (See~\cite{SM} for details). The system displays a period doubling in the magnetization dynamics as observed in several experiments~\cite{Choi2017a,Zhang2017,Randall2021,Frey2022,Xiao2021,Xu2021}.

 The schematic of our protocol is shown in Fig.\ref{fig.sketch.introduction}(a-c); the sensor in the FTC phase is put in contact with an external AC-field of amplitude $h$, and frequency $\omega_h$ along the $z$-direction;
coupling to the sensor through $\hat V= h \sin (\omega_h t +\theta) \sum_{i} \hat{\sigma}_i^z/2$. The sensor coupled to the external field evolves under the  Hamiltonian $\hat H_s+ \hat V$. We focus our discussions on estimating, with maximal accuracy, the amplitude field $h$, nevertheless, the analysis can be directly extended for the estimation of the associated frequency $\omega_h$.
 In all our simulations we use initial separable states, see details in~\cite{SM}, where we also discuss  different initial states.  We average our results over $n_{\rm dis} = 10^{3}$ to $10^{4}$ disorder realizations, depending on the system size.

Optimal sensing can be reached once the period of the kicks in Eq.( \ref{eq.model}) is chosen to be $T \sim \pi /\omega_h$.
 In this case, the spins dynamics is collectively phase-locked with the AC field, allowing a long and coherent interrogation time.  This is a unique trait of FTCs (not present in other non-ergodic phases), and it is essential for the performance of the quantum sensor discussed here.  Moreover, the presence of MBL dynamics due to the disorder, leads to a growth of quantum correlations in the sensor  boosting its performance. This is summarised  schematically in  Fig.\ref{fig.sketch.introduction}(d), where the uncertainty in estimating $h$ decreases with the duration of the sensing protocol.
For the FTC sensor, 
such decrease enhances its perfomance above the usual SQL, $(\Delta h(t))_{SQL} \propto t^{-1}$, displayed by either time-independent many-body Hamiltonians, or time-dependent ones after the transient regime.
Furthermore, such an advantage lasts up to $ \sim {\rm max} [t_{th}, h^{-1}]$, where $t_{th}$ is the thermalization time, which for the FTC sensor increases \textit{exponentially} with its size $N$.

\textbf{\textit{Sensing and quantum Fisher information. } }
 In the frequentist approach to sensing, typically, estimation protocols involve a repetition of the following steps: (i) the initialization of the sensor in an "advantageous/entangled" state; (ii) a time interval in which the sensor interacts with the signal of interest (in our case $h$), so that the unknown parameter is encoded in the state of the sensor; (iii)  a measurement on the quantum sensor.  By collecting the statistics of the repeated protocol, one infers the value of the parameter with maximal accuracy.  Assuming that some prior knowledge of the parameter is known,
the least uncertainty on the estimated parameter is settled by the quantum Cramér-Rao bound (details in~\cite{SM}) $\Delta h(t) \geq 1/\sqrt{ M F_{h}(t) }$
where $F_{h}(t)$ is the quantum Fisher information (QFI) of the probe and $M$ is the number of measurements~\cite{Braunstein1994,Liu2019}.  Notice that for a $N$ many-body probe, the role of repeated measurements $M$ can be played by $N$ if the QFI is additive or can be superlinear in $N$ if correlations are present.
For pure states the QFI has the form,
\begin{equation}
F_{h}(t)/4 =  \langle \psi'(t,h) | \psi'(t,h) \rangle -  
  |\langle \psi(t,h) |\psi'(t,h) \rangle |^2
\end{equation}
with $|\psi'(t,h) \rangle =\partial_{h} |\psi(t,h) \rangle$, the partial derivative w.r.t the estimated parameter.

\begin{figure}
\includegraphics[width = 0.85 \linewidth]{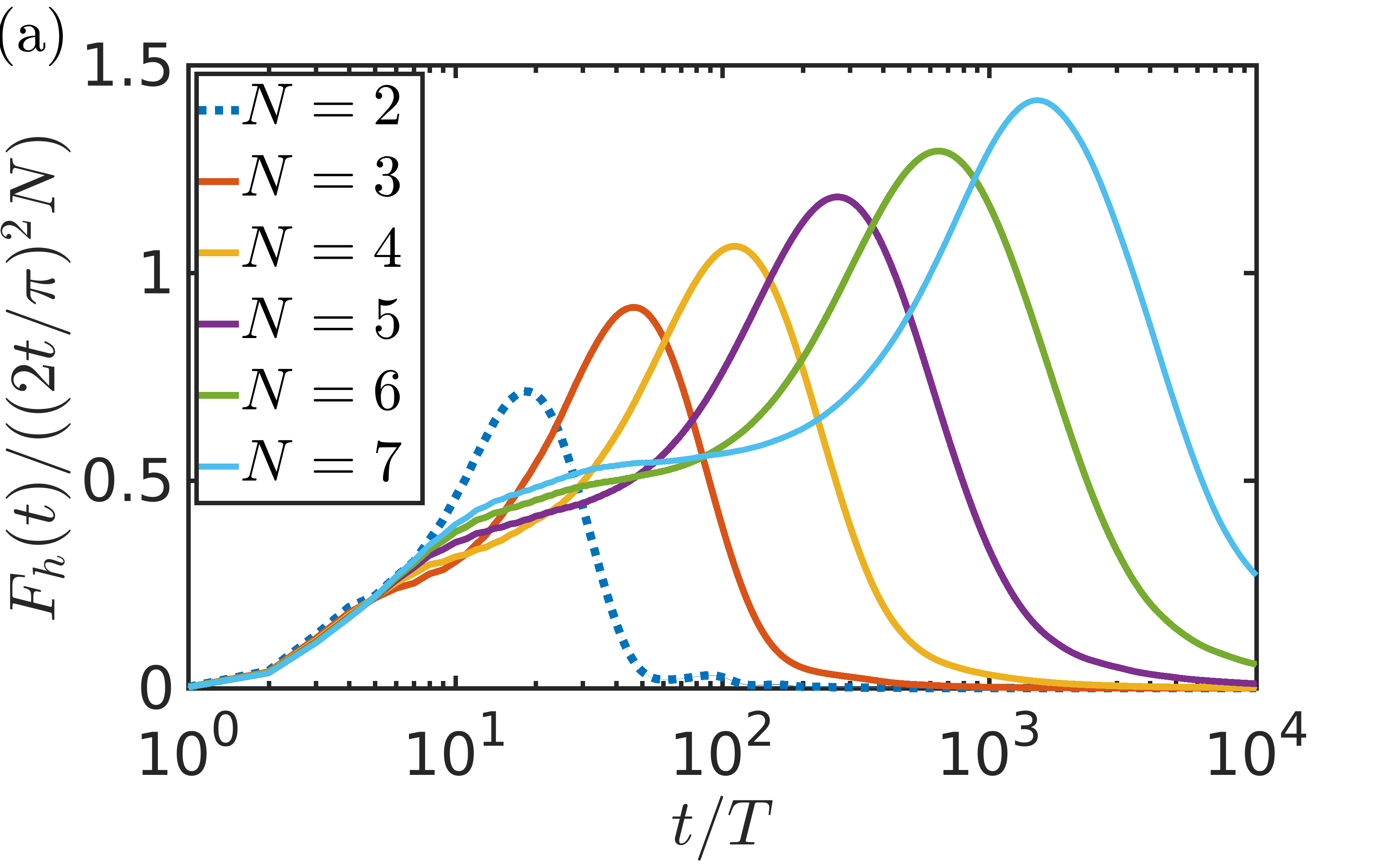}\\
\includegraphics[width = 0.85 \linewidth]{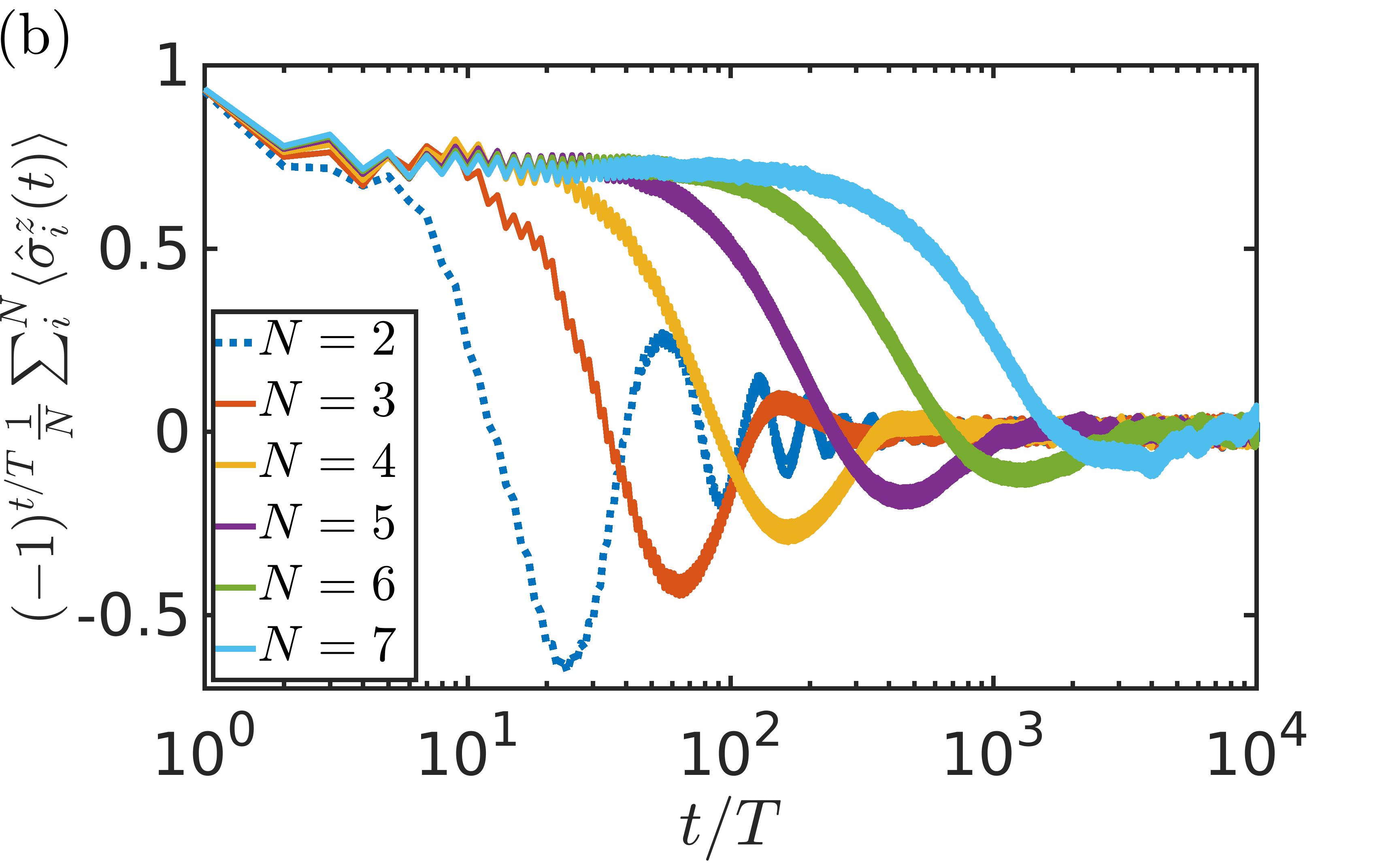}\\
 \caption{ 
  (a) QFI dynamics as a function of the  FTC sensor size $N$ at resonance $\omega_{h} = \pi/T$. After an initial transient, the growth first saturates and then overcomes (due to the many-body correlations increase) the SQL ($F_h(t) \sim t^2$). (b) The corresponding magnetization dynamics.
The parameters here are set to $J=b^z =0.25$, $b^x=0.025$, $\theta = 0$, $\phi =2.8$, $T = 1$, $\omega_{h}=\pi/T$ and $h \rightarrow  0$.
}
 \label{fig.sketch.introduction2} 
\end{figure}

\begin{figure}
\includegraphics[width = 0.85 \linewidth]{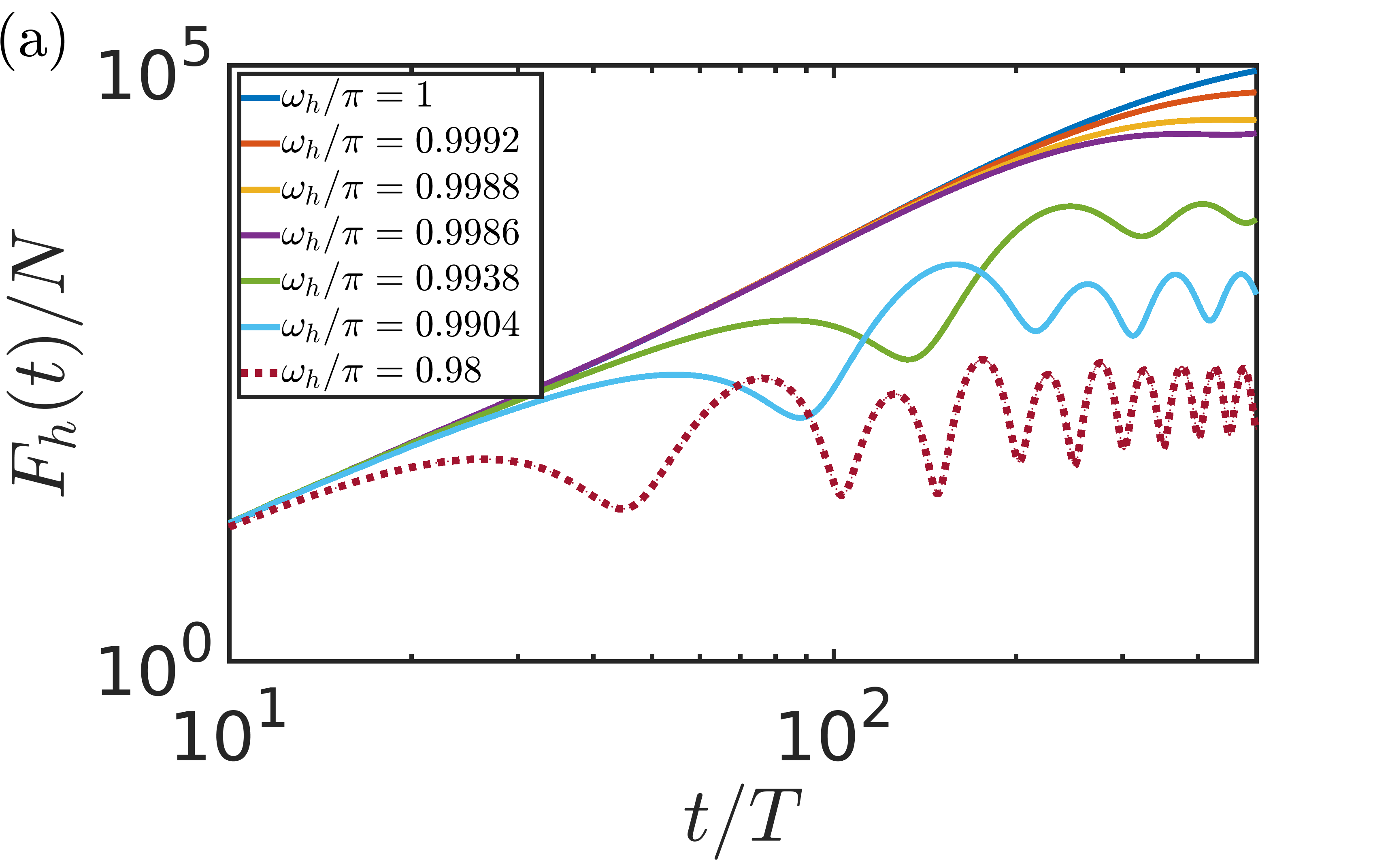}
\includegraphics[width = 0.85 \linewidth]{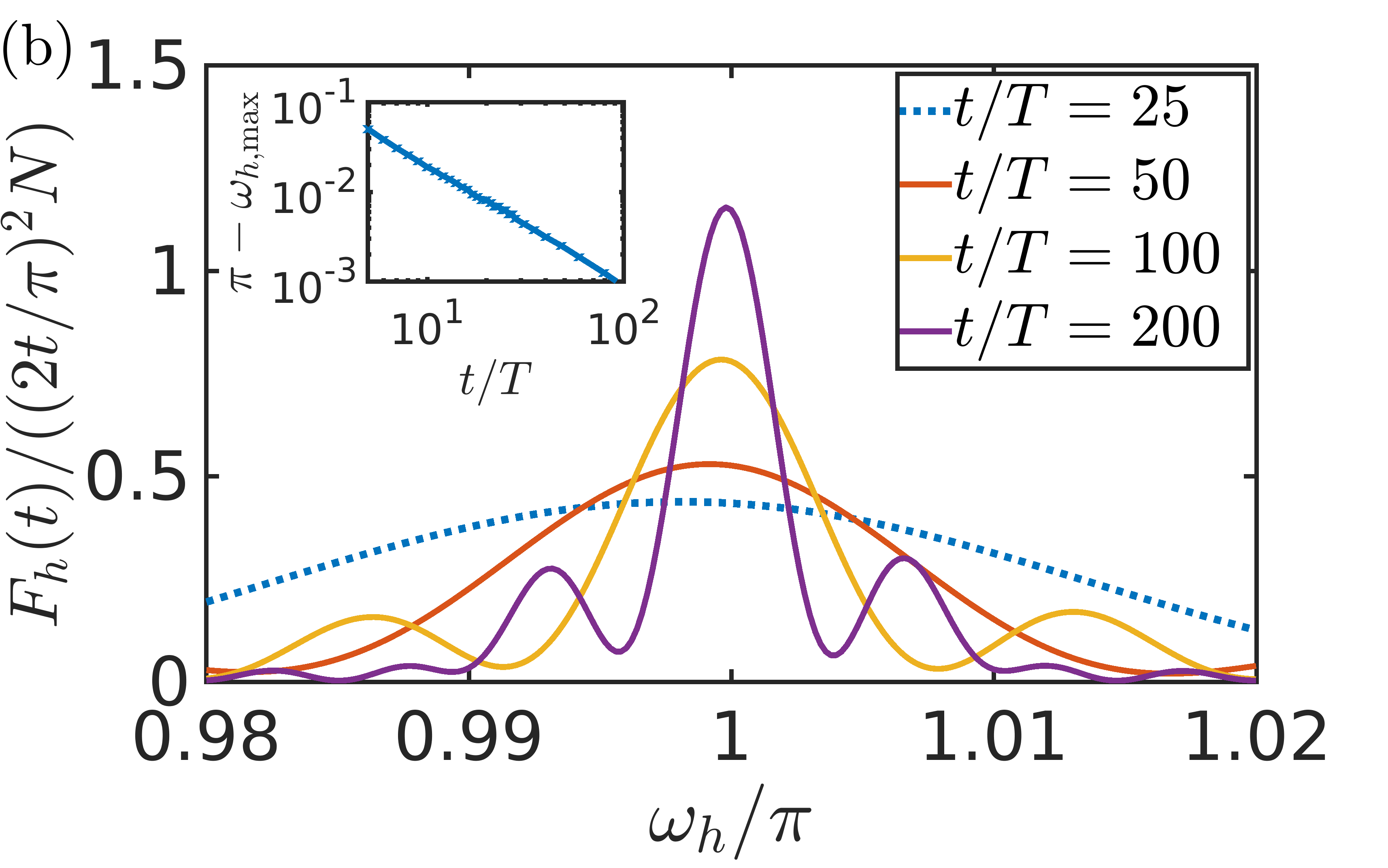}
 \caption{ Linear response for $N=4$, and same parameters as in
Fig.~\ref{fig.sketch.introduction2}, chose to set the sensor in the FTC phase. (a) Dependence of the QFI dynamics with varying the AC-frequency. (b) Frequency dependence of the QFI, at fixed times with a resonant peak emerging at long times. (inset) The position of the peak approaches the FTC frequency with increasing time.
}
 \label{fig.numerical.hac0}
\end{figure}

\textit{\textbf{ Case with $\phi = \pi$, $b_i^x = 0$,  $h \rightarrow 0$}.} 
We consider perfect flips ($\phi = \pi$) during the kicks, no transverse magnetic fields and the \textit{linear response} regime. 
In this case, one can obtain analytical results for the performance of the quantum sensor.   Despite its simplicity, we can grasp several important properties of the quality of the sensing protocol. The QFI reduces to (see\cite{SM})
\begin{equation}
  F_{h \rightarrow 0}(t) = (\varphi(t))^2 \rm{var}(\sum_{i} \hat{\sigma}_i^z).
\end{equation}
  Correlations among the spins, encoded in the variance of the total spin, may provide an enhanced sensing for the system (with a possible super-linear scaling with $N$). The
time-dependent prefactor $\varphi(t) \equiv \int_0^t \sin (\omega_h t' +\theta)  (-1)^{[t'/T]} dt'$  is the phase accumulated during the dynamics.
This term is the same as for non-interacting sensing protocols, such as Carr-Purcell pulse trains~\cite{Carr1954} or periodic dynamical decoupling sequences~\cite{Khodjasteh2005}. 
 The maximun accumulated phase (when the spin dynamics is  in phase with the AC-field)
leads to a quadratic growth with time ($\sim t^2$).
This happens when the spins are in resonance with the field, $T = \pi/ \omega_{h}$, which for stroboscopic times $t=pT$ reduces to $  \varphi (pT)  \to  (2pT/\pi) \cos\theta $ 
 with $p$ an integer number.
Slightly deviations $O(\epsilon)$ from the resonant case 
lead to off-resonance correction at time scales at the order of $t_{\rm{off}}/T = \theta  /\epsilon$, for which the growth of accumulated phase slows down.  In order to see a scaling with an exponent $\beta > 1$ (see Fig.\ref{fig.sketch.introduction}(d)), we have to consider the generic case discussed in the next paragraph.

 \textit{\textbf{General case. }} We now turn to the more realistic scenario, considering both imperfections in the kicks ($\phi \neq \pi$) as well as  the presence of a transverse field, $b_i^x \neq 0$, that leads to decoherence in the dynamics
 The interactions among the spins now strongly impact the dynamics of the system. 
In the resonant case  ($T \sim \pi/\omega_h$), and in the linear response regime, the dynamical behaviour of the QFI is displayed in Fig.(\ref{fig.sketch.introduction2}a) for different sensor sizes $N$. In all cases, after an initial transient growth, the QFI reaches first a plateau, saturating the SQL growth in time ($\sim t^2$).
At later times, the QFI overcomes the SQL due to correlation growth till the system thermalizes, at times exponentially divergent with $N$.  In this intermediate regime, the maximun QFI has a higher than quadratic growth in time $F_{h}(t)/N \sim t^{2 \beta(t)}$ with $\beta(t) > 1$. Similarly, the correlations along the AC field direction, quantified by the spin variance, shows an initial transient growth followed by a stable plateau, and a subsequent growth towards system thermalization (see \cite{SM}).
 The different regimes reflect the FCT dynamics,  as shown in Fig.(\ref{fig.sketch.introduction2}b) for the total magnetization.
  After an initial transient time, $\mathcal{O}(T)$, it reaches a plateau whose lifetime corresponds to that of the QFI plateau. For longer times, dephasing due to MBL leads the magnetization to slowly decay towards its thermal value. The thermalization time (exponentially large in $N$) agrees with the time window for which the QFI overcomes the SQL scaling (i.e., the time when the QFI reaches its peak in Fig.~\ref{fig.sketch.introduction2}(a).
 
 Indeed, Fig.\ref{fig.sketch.introduction2} summarizes the potential of the FCT as a quantum sensor, it beats the SQL for time intervals that scale exponentially with the size of the sensor.
 
The optimal performance crucially depends on the tuning of the sensor frequency with the AC-field, as  shown in Fig.\ref{fig.numerical.hac0}(a-b).
 Their mismatch sets a new characteristic time that scales with a power of the detuning.
The response of the sensor to the  AC- field, has nevertheless a well structured form, as shown in Fig.~\ref{fig.numerical.hac0}(b). The  collective interactions among the spins stabilize the sensor against noise. 

As discussed previously, the performance of the FTC sensor can be enhanced by its initial preparation.  For low-entangled initial states, similarly to the above case, there is $\rm{var}(\sum_{i} \hat{\sigma}_i^z)$ factor that can lead to super-extensive (with $N$) behaviour. High-entangled GHZ-like initial states, known to be fragile to decoherence, show instead a different behavior with a continuous but very slow decay of the QFI dynamics - see \cite{SM}. In this case, despite the deterioration in performance, it is still attractive to exploit the sensor in a beneficial way due to such slow relaxation dynamics.
  
\begin{figure}
\includegraphics[width = 0.85 \linewidth]{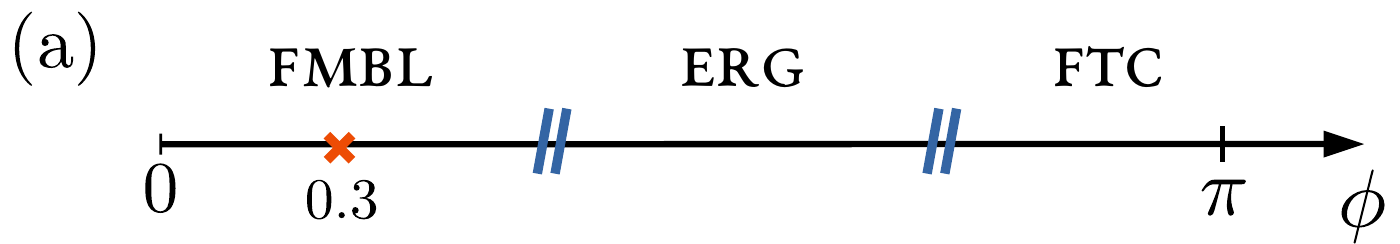}
\includegraphics[width = 0.9 \linewidth]{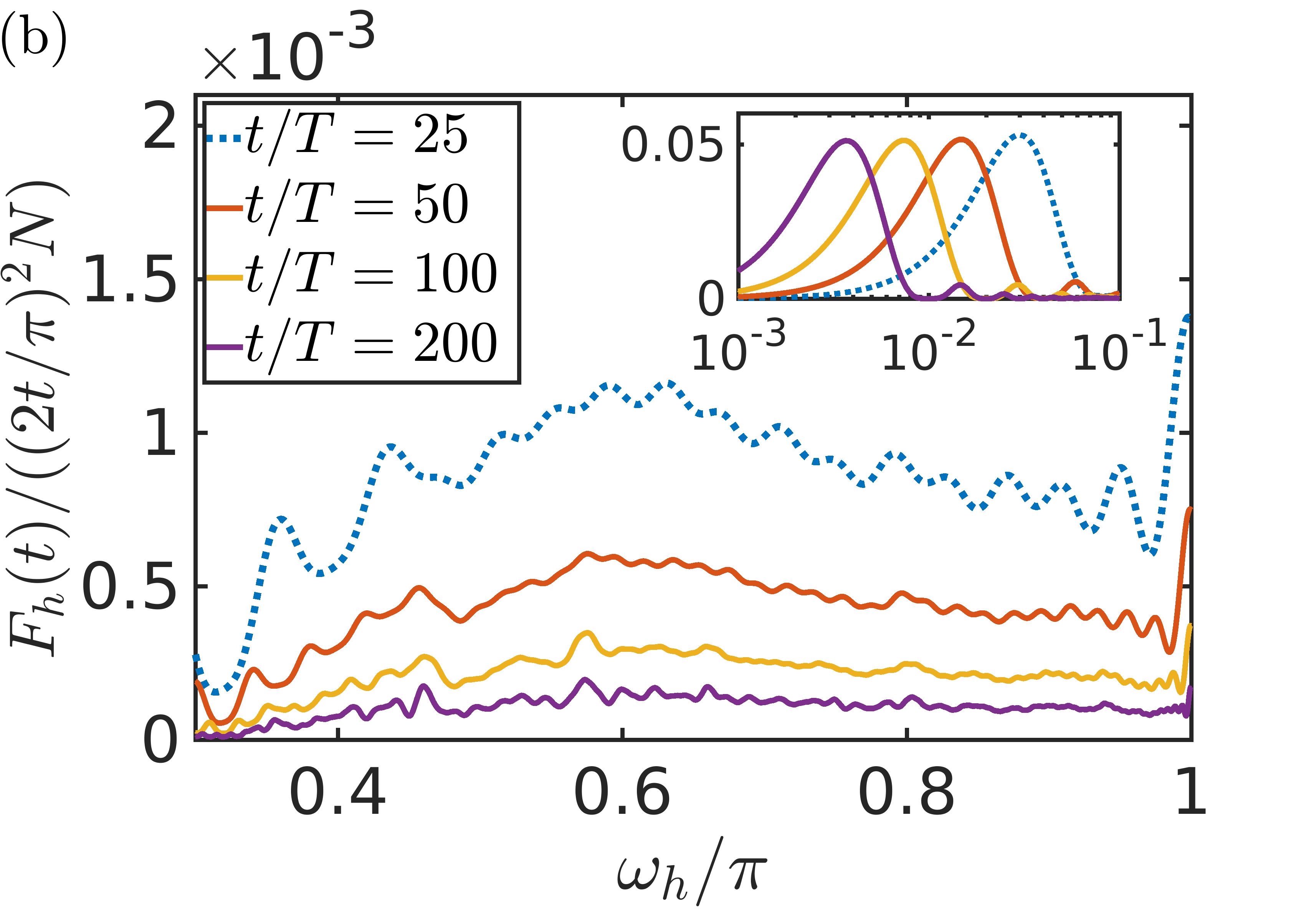}
 \caption{ (a) Schematic phase diagram of the sensor as a function of the global $\phi\in [0,\pi]$ rotation (induced from the periodic kicks), going from a Floquet MBL phase (FMBL), to an ergodic phase (ERG) and a Floquet time crystal (FTC). (b) The performance of the sensor in FMBL phase, displayed by the QFI, for parameters
 $\phi = 0.3$ (red cross in the schematic picture),  $J=b^z =0.25$, $b^x=0.025$, $\theta = 0$,  $T = 1$ and $N=4$. The (noisy) response is upper bounded by the SQL. For decreasing frequencies $\omega_h \rightarrow 0$ (i.e. DC limit) the QFI displays the $t^2$ growth (peaks in the inset panel) due to the localization dynamics. The performance in
the ergodic phase is similar to the FMBL case~\cite{SM}.
}
 \label{fig.numerical.ergodic.g0}
\end{figure}

 It is important to understand to which extent the properties of the sensor depend on the existence of the TC. This crucial point is reported in Fig.~\ref{fig.numerical.ergodic.g0} where, by changing the global $\phi$ rotation, we can access to an MBL phase without time cristalline order, that we denote by (FMBL), and to an ergordic phase (ERG), as schematically depicted in Fig.~\ref{fig.numerical.ergodic.g0}(a). The QFI in the FMBL phase is displayed in Fig.~\ref{fig.numerical.ergodic.g0}(b). Due to the absence of the phase-locked resonant features with the AC-field,  decisive for the performance of the sensor, the QFI  in the FMBL can never beat neither reach the SQL. In the ergodic phase, the thermalization time is of the order of a few kicks, roughly independent on the system size. Also in this case, the QFI has no
 structured response to the probe field and its growth in time is always upper bounded by the SQL, as shown in \cite{SM}.
 
\begin{figure}
\includegraphics[width = 0.85 \linewidth]{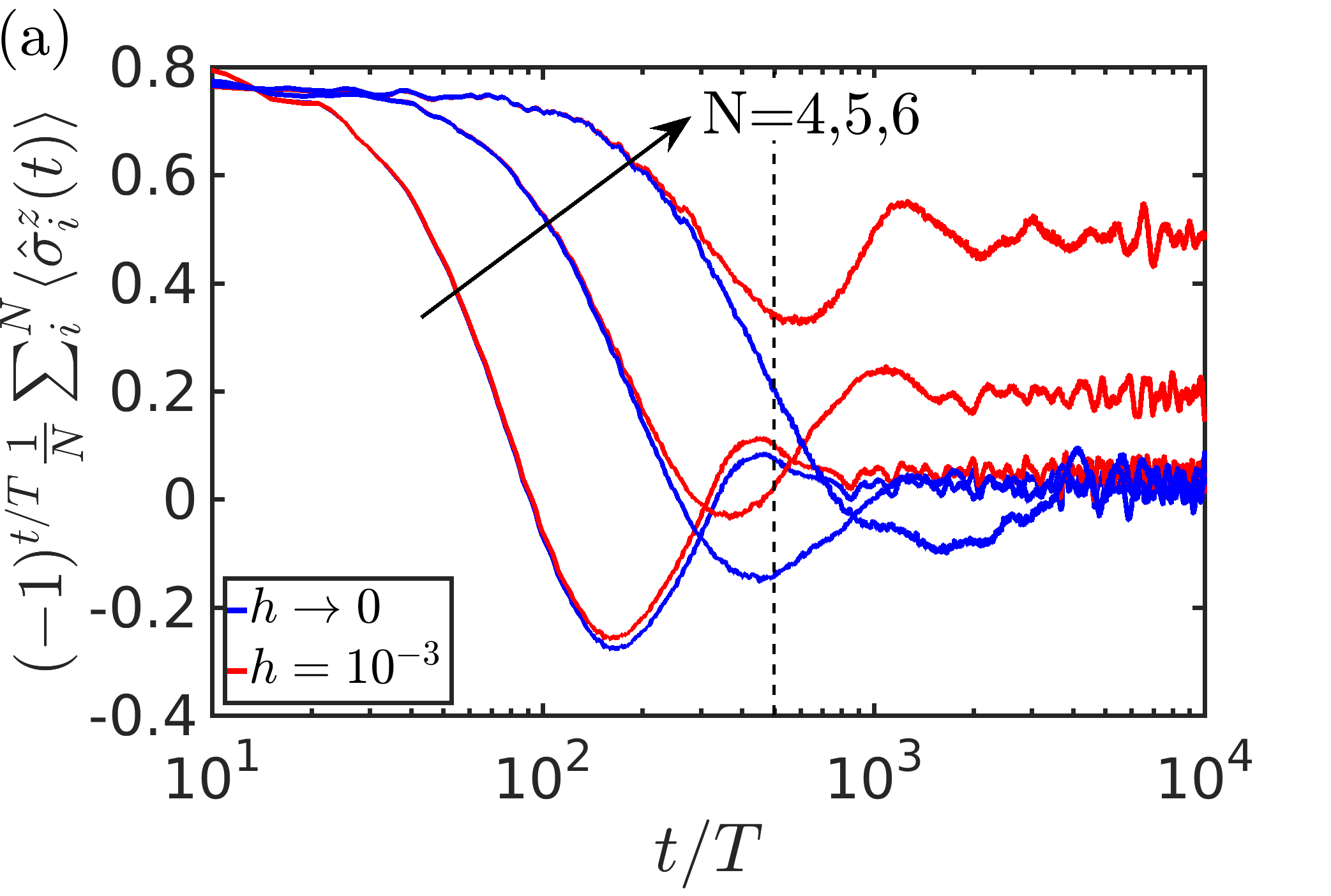}
 \includegraphics[width = 0.85 \linewidth]{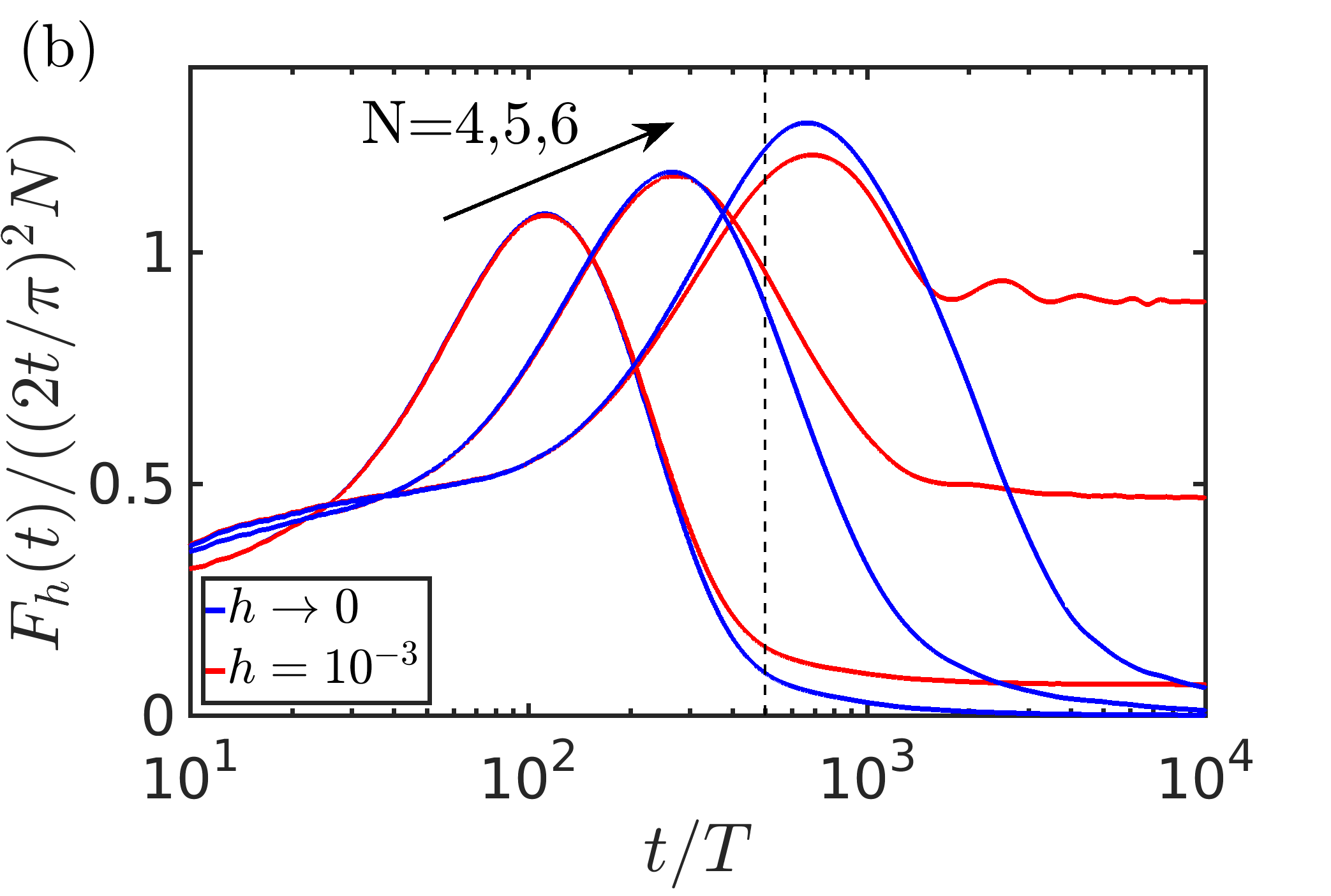}
 \caption{ Performance of the FTC sensor for finite field amplitudes $h$. The parameters are chosen to set the
sensor in the FTC phase (same as in Fig.~\ref{fig.sketch.introduction2} but for varying $h$
field). (a) Magnetization dynamics, with dominant AC-effects
appearing at time scales of the order of $t_{AC} \sim h^{-1}$. (b) The
QFI dynamics crosses over to a quadratic growth in time for
$t \gtrsim t_{AC}$. The vertical dotted line is a guide to the eye, setting the time scale $t_{AC}$.
 }
 \label{fig.numerical.hacneq0.varhac}
\end{figure}
 
 Finally, we consider finite field amplitudes $h$, beyond linear response. The AC-field becomes dominant for sufficiently long times and the sensor state is strongly influenced by the external field (therefore we expect a high sensibility for its estimation). Such effects appear  at time scales of the order of $t_{ac} \sim h^{-1}$. In the resonant case, the AC-field tends to stabilize the period-doubling magnetization of the system, as shown in Fig.~\ref{fig.numerical.hacneq0.varhac}(a). In this regime the QFI  ($t  \gtrsim t_{ac}$) shows a crossover to a purely quadratic growth in time - Fig.~\ref{fig.numerical.hacneq0.varhac}(b)-  thus recovering the SQL. This behavior is apparently generic, i.e., independent on the sensor characteristics~\cite{SM}. Therefore, one may interpret such time scale, $t_{ac}$, as the strong AC-field regime, where the detailed many-body properties of the sensor become unimportant for the measurement estimation.
 As in linear response, off-resonant effects appear at times that scale with a power of the detuning, suppressing the period-doubling dynamics. The QFI slows down its growth recovering its quadratic growth. Also in this case, the  QFI keeps its structured  dependence with the  probe, with a peak at resonance~\cite{SM}.

 \textit{\textbf{Conclusions. }} In this work we discuss FTCs as quantum sensors for AC-fields. Their optimal performance - reached once the sensor is set on period-doubling resonance to the field - is shown to offer several advantages, overcoming the shot-noise limit, allowing long-time sensing measurements times exponentially large with the number of spins, and being inherently robust to noise or imperfections in the protocol. The sensor offers, moreover, a promising dual role as a probe for unveiling the underlying ordering of the system in general cases. It would be interesting to explore other forms of time crystal sensors in closed, see e.g.~\cite{Surace2019,Giergiel2018,Nurwantoro2019,Giergiel2020,Pizzi2021,Giachetti2023} or in open systems with either discrete~\cite{Gong2018} or continuous~\cite{Iemini2018,Buca2019} time translation symmetry breaking.

\textit{\textbf{Acknowledgements. }} F.I. acknowledges financial support from Alexander von Humboldt foundation and the Brazilian funding agencies CAPES, CNPQ, and FAPERJ (Grants No. 308205/2019-7, No. E-26/211.318/2019, No. 151064/2022-9, and No. E-26/201.365/2022), and by the Serrapilheira Institute (grant number Serra – 2211-42166).  A.S. acknowledges financial support from the Spanish Agencia Estatal de Investigacion, Grant No. PID2022-141283NB-I00, the European Commission QuantERA grant ExTRaQT (Spanish MICINN project PCI2022-132965), by Ministerio de Ciencia e Innovación with funding from European Union NextGenerationEU(PRTR-C17.I1) and by
Generalitat de Catalunya and the Spanish Ministry of Economic Affairs and Digital Transformation through the QUANTUM ENIA project call – Quantum Spain project, and the European Union through the Recovery, Transformation and Resilience Plan – NextGenerationEU within the framework of the Digital Spain 2025 Agend.
  R.F. ackowledges financial support from PNRR MUR project PE0000023-NQSTI and by the European Union (ERC, RAVE, 101053159). Views and opinions expressed are however those of the author(s) only and do not necessarily reflect those of the European Union or the European Research Council. Neither the European  Union nor the granting authority can be held responsible for them.


%

\widetext
\clearpage

\begin{center}
	\large \textbf{Supplemental Material} \\ \vspace{0.3cm}
	Fernando Iemini, Rosario Fazio and Anna Sanpera
\end{center}
\setcounter{section}{0}
\setcounter{equation}{0}
\setcounter{figure}{0}
\setcounter{table}{0}
\setcounter{page}{1}
\makeatletter
\renewcommand{\theequation}{S\arabic{equation}}
\renewcommand{\thefigure}{S\arabic{figure}}
\renewcommand{\bibnumfmt}[1]{[S#1]}
\renewcommand{\citenumfont}[1]{S#1}

In this Supplemental Material we give further details on the computation of the QFI, the sensor phase diagram and its performance once tuned along its FTC, ergodic and FMBL phases.

\section{Computation of the QFI}
As discussed in the main text, for pure states the Fisher information is defined as,
 \begin{eqnarray}
 F_{h}(t)/4 = \langle  \psi'(t,h) | \psi'(t,h) \rangle -  |\langle \psi(t,h) | \psi'(t,h) \rangle |^2
\end{eqnarray}
where $|\psi'(t,h) \rangle = | \partial_h \psi(t,h) \rangle$ the partial derivative with respect to the estimated parameter.

One can derive a simpler form to compute the Fisher information, reformulating it in terms of time-correlated variances, as follows.
 The unitary operator for the time evolution
 $ \hat U(t) = \mathcal{T}e^{\int_0^t -i \hat H(t')dt'}$ with $\mathcal{T}$ the time ordering operator, can be expanded for infinitesimal time steps as
 \begin{equation}
  \hat U(t) = \lim_{K \rightarrow \infty} \prod_{k=1}^K e^{-i \hat H(t_k) dt} = \lim_{K \rightarrow \infty} \prod_{k=1}^K \hat u_k
 \end{equation}
 with $dt = t/K$, $t_k = k dt$ and $\hat u_k \equiv e^{-i H(t_k) dt}$. The partial derivative of the Floquet unitary follows directly
 \begin{eqnarray}
  \frac{\partial \hat U(t)}{\partial h_{\rm ac}} &=& \lim_{K \rightarrow \infty} \sum_{k'=1}^K \left( \prod_{k>k'} \hat u_{k} \right)
  \frac{\partial \hat u_{k'}}{\partial h_{\rm ac}}
  \left( \prod_{k< k'} \hat u_k \right).
 \end{eqnarray}
 and can be rewritten in the simpler form
 $\frac{\partial \hat U(t)}{\partial h_{\rm ac}} = \hat U(t)
 \hat S(t)$, with
\begin{eqnarray}\label{eq.Smatrix.commuting}
\hat S(t) &=& -i \int_0^t \hat U(t')^\dagger
(\partial_h \hat V(t'))
\hat U(t') dt' \\
&=&  -i \int_0^t \hat U(t')^\dagger
(\sin(\omega_h t' + \theta) \sum_i \frac{\hat \sigma_i^z}{2} )
\hat U(t') dt'
\end{eqnarray}
the scrambled signal operator. Therefore, from this formulation the Fisher information is given by,
\begin{equation}
 F_{h}(t)/4 = \rm{var}(\hat S(t))
\end{equation}
where $\rm{var}(\hat A) = \langle \hat A^2 \rangle - \langle \hat A \rangle^2 $ is the variance of the operator with the expectation value computed for the initial state of the sensor $\langle ... \rangle \equiv \langle \psi(0)| ...| \psi(0) \rangle $.

 The standart quantum limit can be derived from the above relation. Considering a single spin model, the maximun quantum Fisher information is bounded by the maximun variance of the scrambled signal operator of Eq.\eqref{eq.Smatrix.commuting}, which for integer periods of time corresponds to $(2t/\pi)^2$. The SQL is therefore the quantum Fisher information bound for $N$ non-interacting spins, which due to the additivity of the Fisher information corresponds to $N (2t/\pi)^2$.

\section*{Phase Diagram with no AC field}

We discuss in this section the phase diagram of our model in the case of no AC fields ($h =0$), expanding  the phenomenology of the Floquet Time Crystals, Ergodic and Floquet Many-Body Localized phases.

\begin{figure}
 \includegraphics[scale = 0.25]{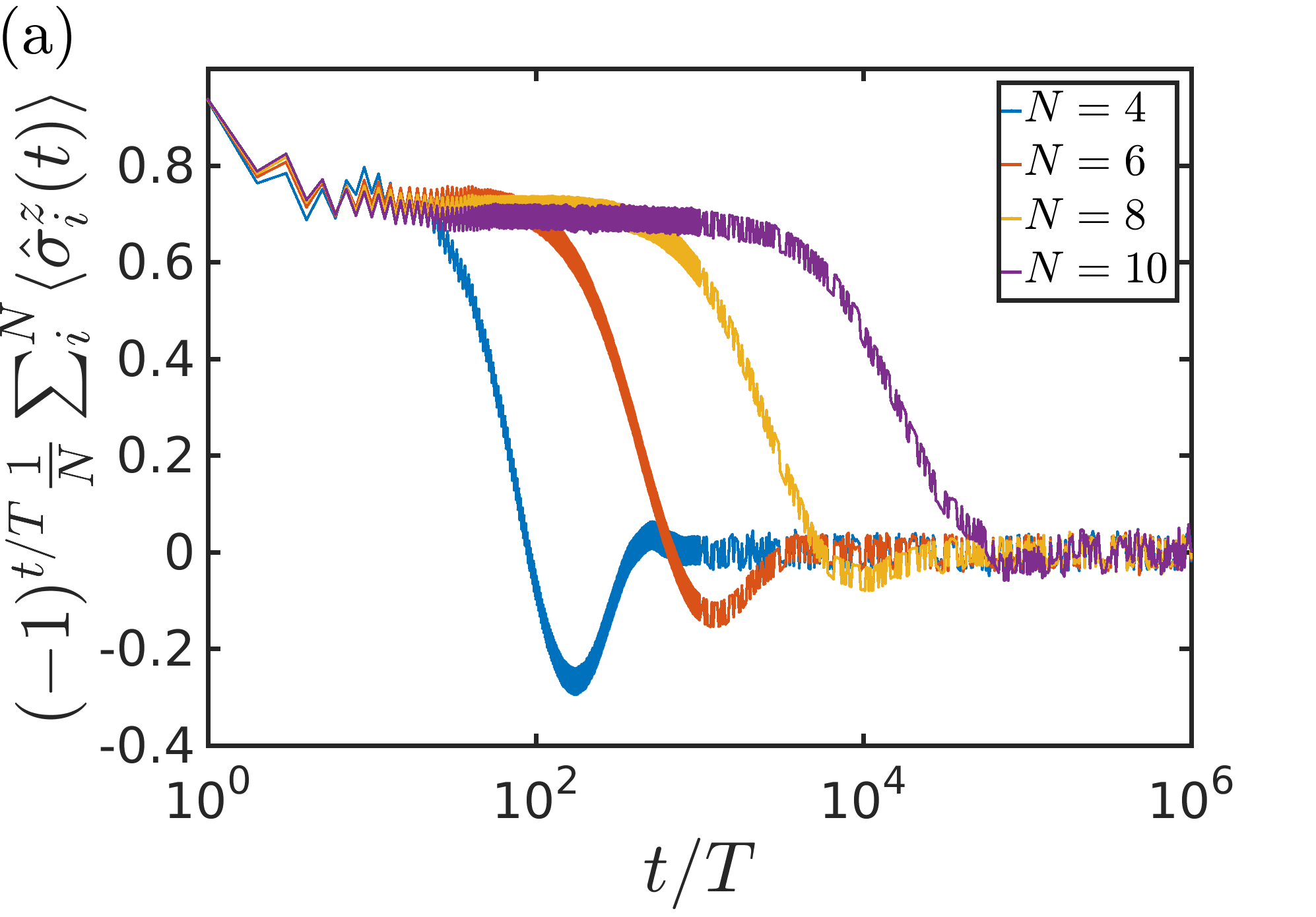}
  \includegraphics[scale = 0.25]{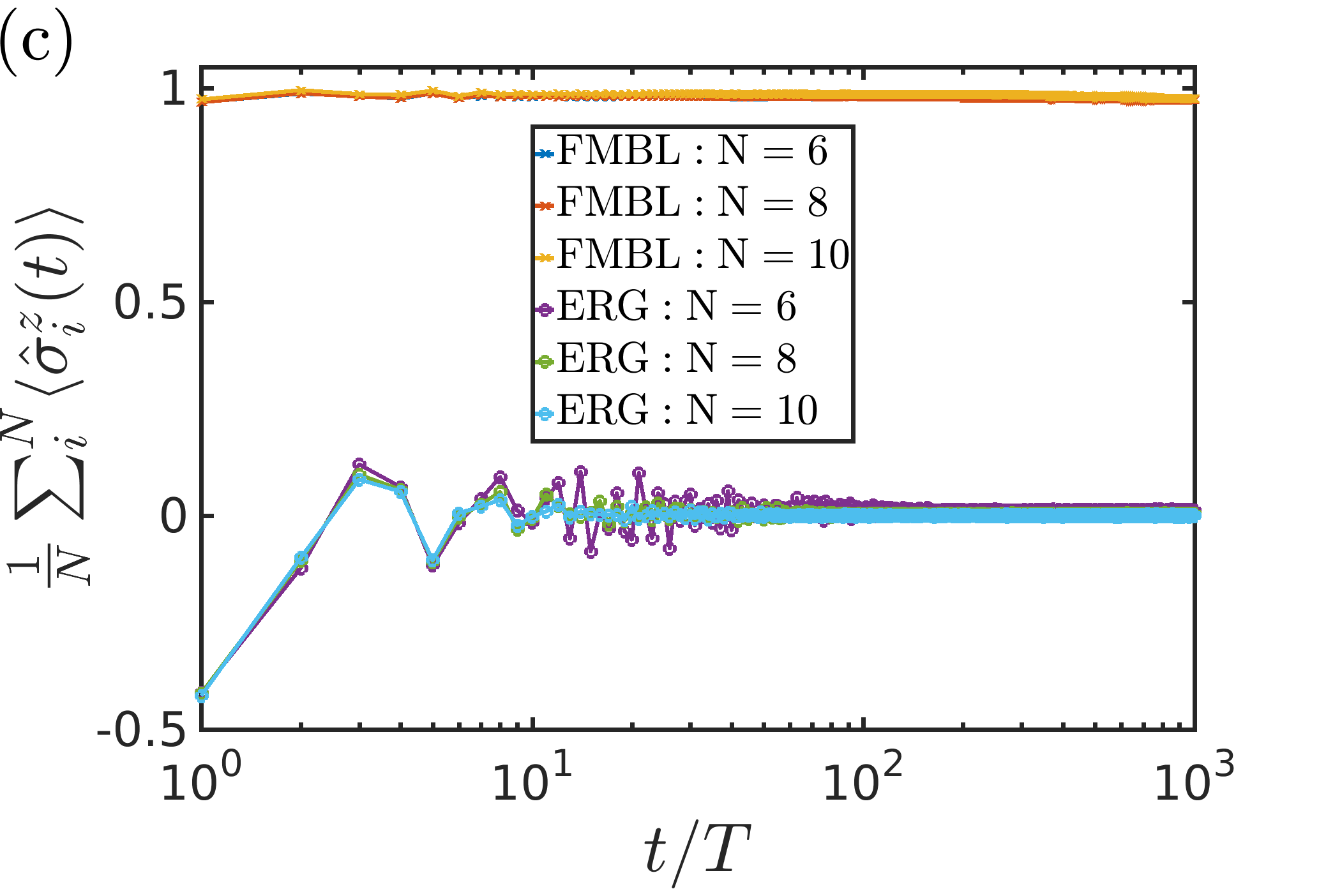}
  \includegraphics[scale = 0.25]{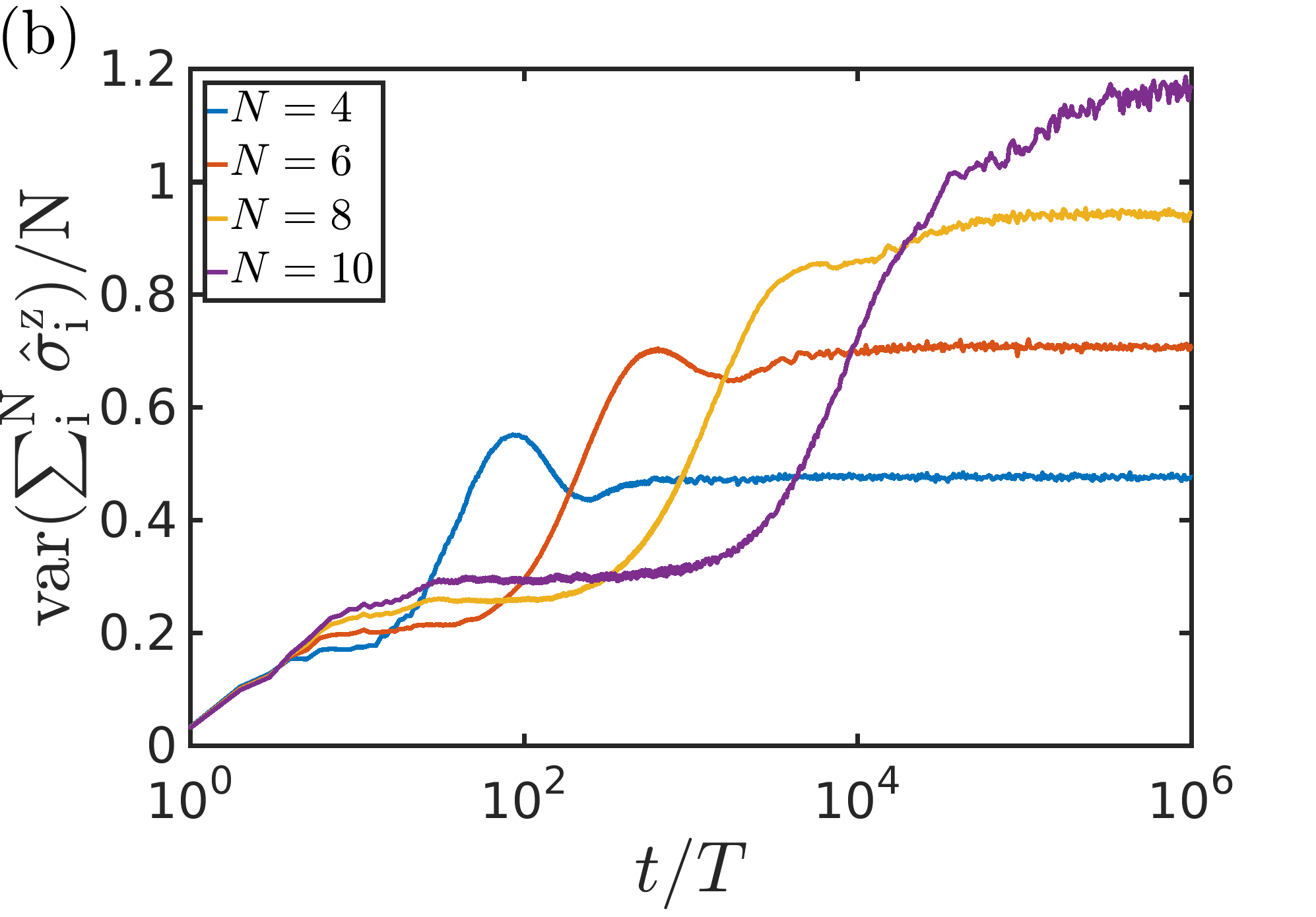}
  \includegraphics[scale = 0.25]{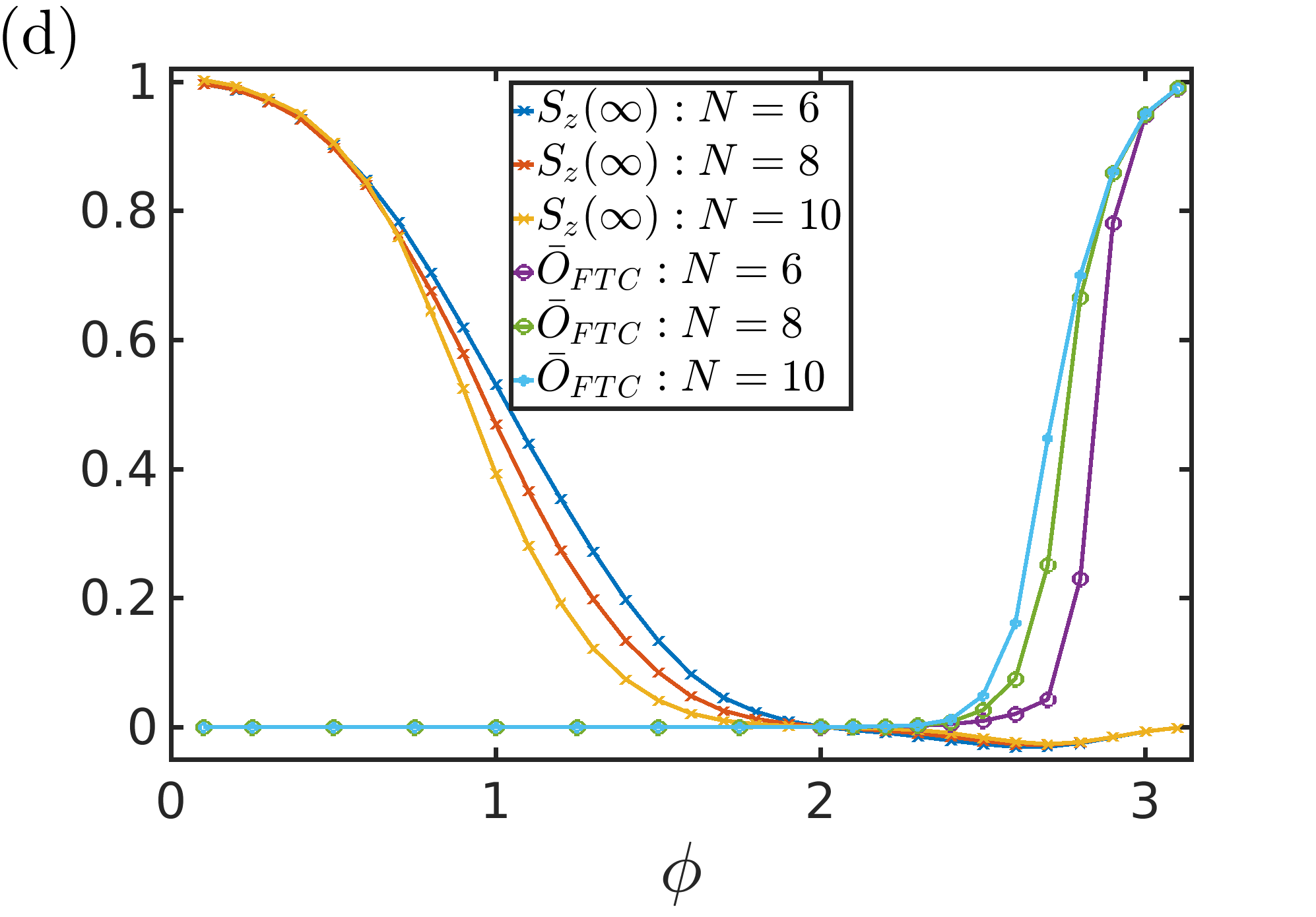}
  \includegraphics[scale = 0.25]{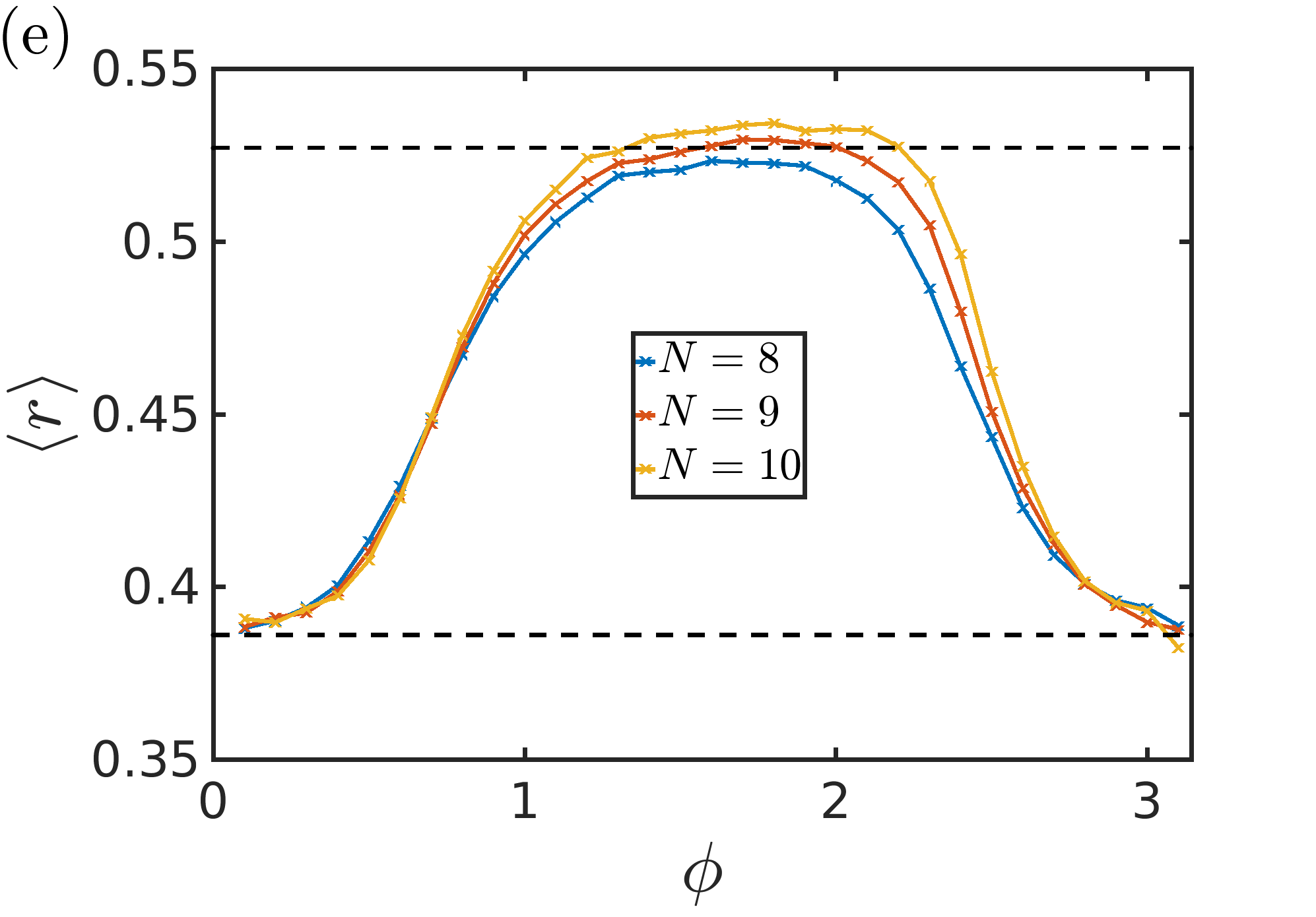}
 \caption{  Phase Diagram for the model with no AC field ($h=0$) with parameters $J=b^z =0.25$, $b^x=0.025$, $\theta = 0$, $T = 1$, $\omega_{h}=\pi/T$ and varying $\phi$ phase.
 Focusing first in the FTC phase, with $\phi = 2.8$, in \textbf{(a)} we show its period doubling dynamics which after a plateau decays to its thermal value.
 In \textbf{(b)} we show the corresponding dynamics for the spin variance along the z-direction.
 In both cases we see a full thermalization at times scales exponentially with system size.
 Turning to ergodic ($\phi = 2$) and FMBL ($\phi = 0.25$) phases,  we show in \textbf{(c)} their magnetization dynamics.
 In \textbf{(d)} we show the infinity average magnetization, $S_z(\infty) = (1/N)\sum_i^N \langle \hat \sigma^z_i(\infty) \rangle$, and period doubling averaging, $\bar{O}_{\rm FTC} = (1/K) \sum_{n=1}^{K} (1/N)(-1)^{t/T} \sum_i^N \langle \hat \sigma^z_i(t) \rangle$ for $K = 10^3$ periods.
 Panel \textbf{(e)} shows the ratio $\langle r \rangle$ for the level-spacing statistics of the model, with the dashed lines indicating $\langle r \rangle \sim 0.386$ and
 $\langle r \rangle \sim 0.527$, for Poisson and Wigner-Dyson distributions, respectively.
 All results are averaged over $\approx 10^4 \,(N=4),\, 5000 \, (N=6,8,9),\,  500 \,(N=10)$ disorder realizations.
 }
 \label{fig.phase.diagram.noAC}
\end{figure}

We show our results in Fig.\eqref{fig.phase.diagram.noAC}.
Once the system is tuned into its FTC phase the magnetization displays, after an initial short transient time of the order $\mathcal{O}(t/T)$, a stable plateau for its period doubling dynamics.  For longer times the period doubling  slowly decay with time to its thermal value due to dephasing among the spins - see Fig.(\ref{fig.phase.diagram.noAC}a).
The dynamics of the correlations along the z-direction, here quantified by the spin variance, shows similar timescales  - Fig.(\ref{fig.phase.diagram.noAC}b). After an initial  transient growth it reaches a plateau which lasts till the magnetization plateau. Later it restarts its growth till a full thermalization of the system.
The thermalization time for both magnetization and correlations agree, and moreover are exponentially large with system size.
In contrast, the magnetization in the ergodic phase shows a fast relaxation for a time scale roughly independent on the system size, while in the FMBL phase the magnetization fails to thermalize even for infinitely long times, as shown in Fig.(\ref{fig.phase.diagram.noAC}c).

In order to derive a phase diagram of the model we can study the average magnetization of the system dynamics and its level spacing distributions.
 In Fig.(\ref{fig.phase.diagram.noAC}d) we show the infinity time magnetization and average period doubling for varying $\phi$ phase. While for small phases the system fails to thermalize with a nonull infinity time magnetization, characterizing a FMBL phase, increasing the phase we see a thermalization and the appearance of an ergodic phase. For larger phases $\phi \sim \pi$ we observe the  appearance of nonnull period doubling magnetization indicating a FTC phase.
 In order to distinguish the various phases we can also study the level spacing distributions of the system, exploring the properties of the Floquet unitary and its eigenvalues.  Computing the consecutive quasienergy gaps of the Floquet unitary, $\delta_i = e_{i+1} - e_i$,
  its level statistics can be characterized by the ratio
  $\langle r \rangle = (1/d) \sum_i^d  \min(\delta_i,\delta_{i+1})/\max(\delta_i,\delta_{i+1}) $, with $d+1$ the dimension of the Hilbert space. While nonergodic phases are expected to have distributions closer to Poisson, with $\langle r \rangle  \sim 0.386$, ergodic ones follow a Wigner-Dyson distribution, with $\langle r \rangle \sim 0.527$.
  We show in Fig.(\ref{fig.phase.diagram.noAC}e) the ratio of the model
 for varying phase $\phi$. As expected, for small or closer to $\pi$ phases we have nonergodic distributions, while in the intermediate values an ergodic phase with Wigner-Dyson distributions.

\section*{FTC sensor}

We discuss in this section further details on the performance of the sensor in its FTC phase for finite amplitude fields.

\begin{figure*}
\includegraphics[width = 0.45 \linewidth]{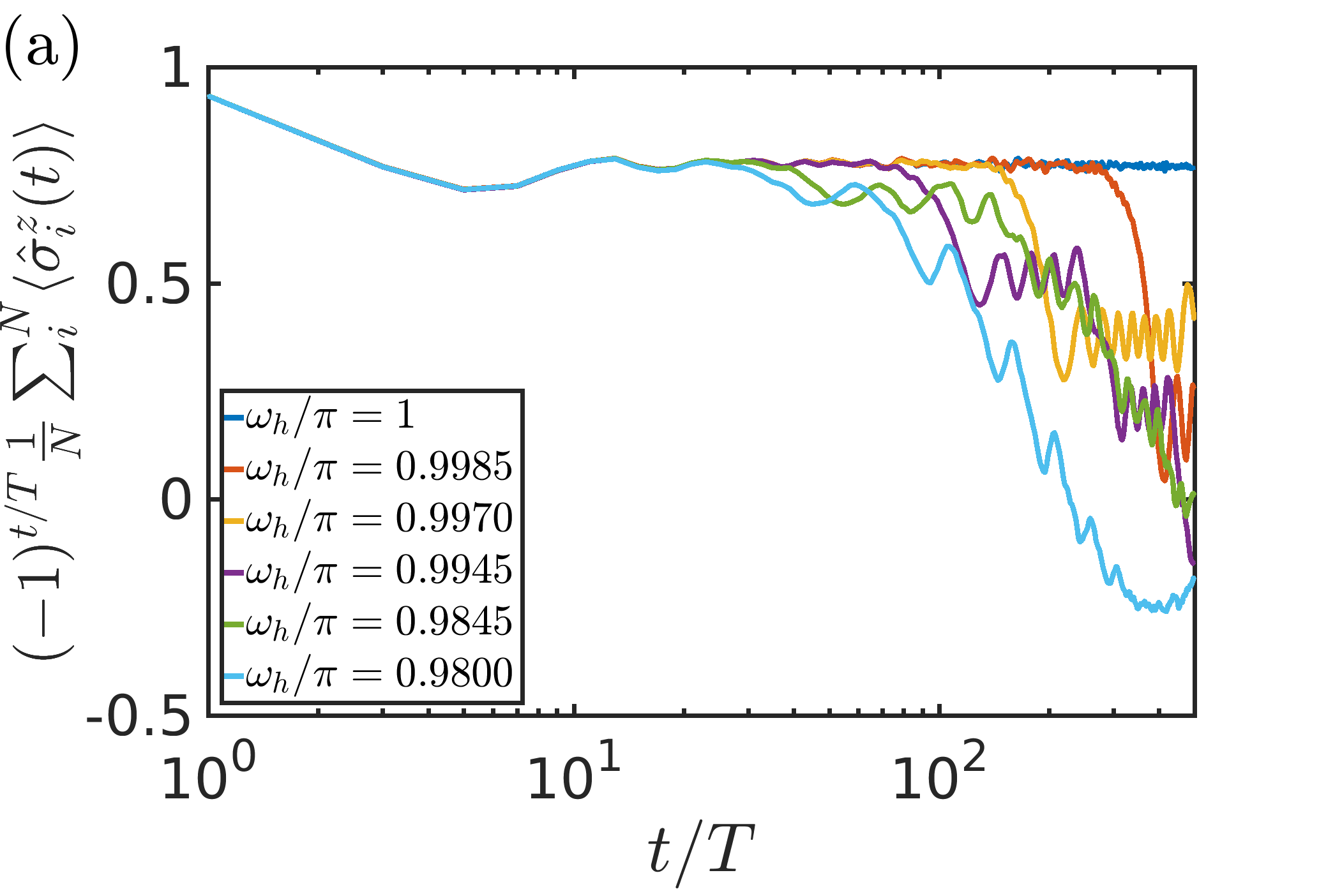}
\includegraphics[width = 0.45 \linewidth]{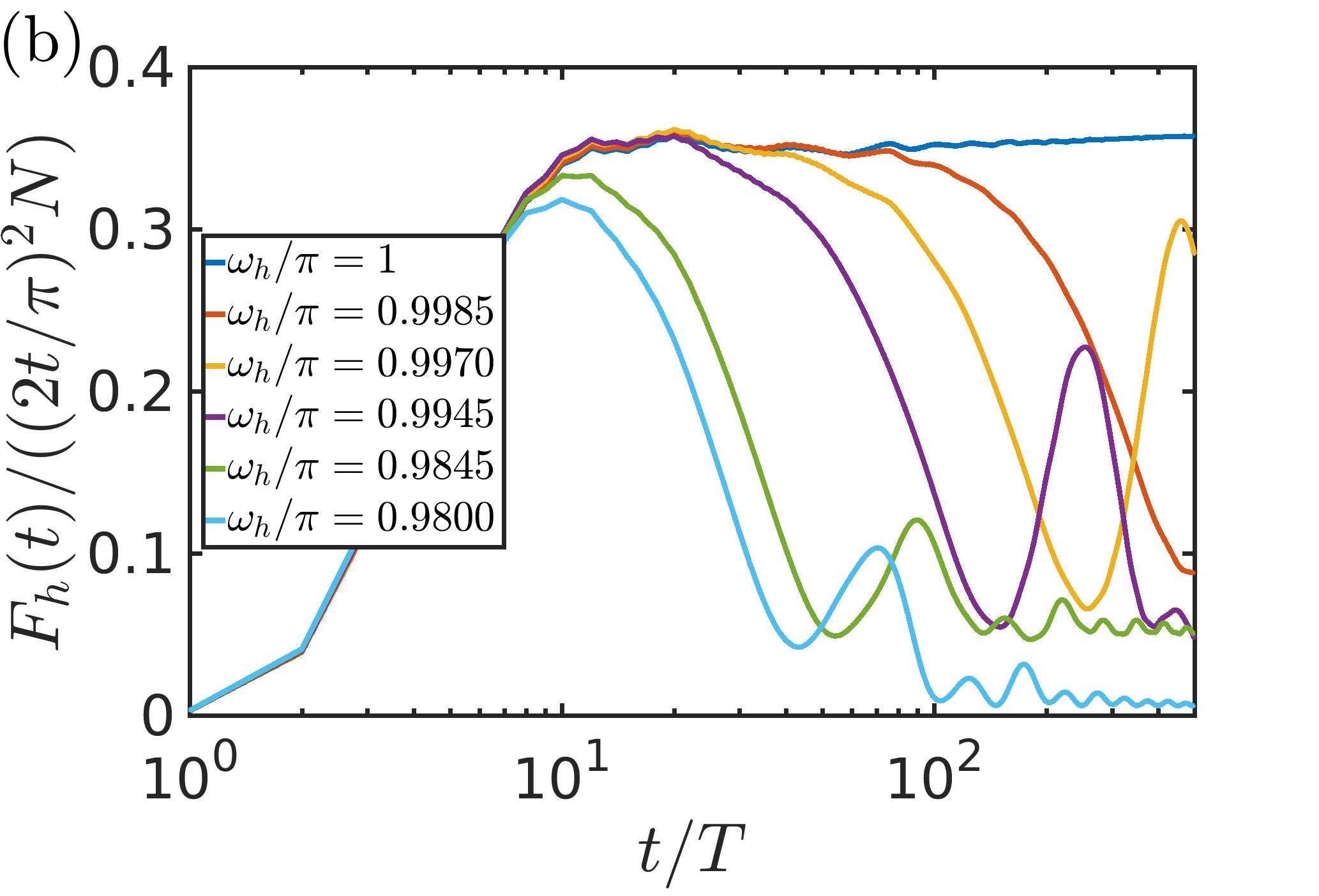}
\includegraphics[width = 0.45 \linewidth]{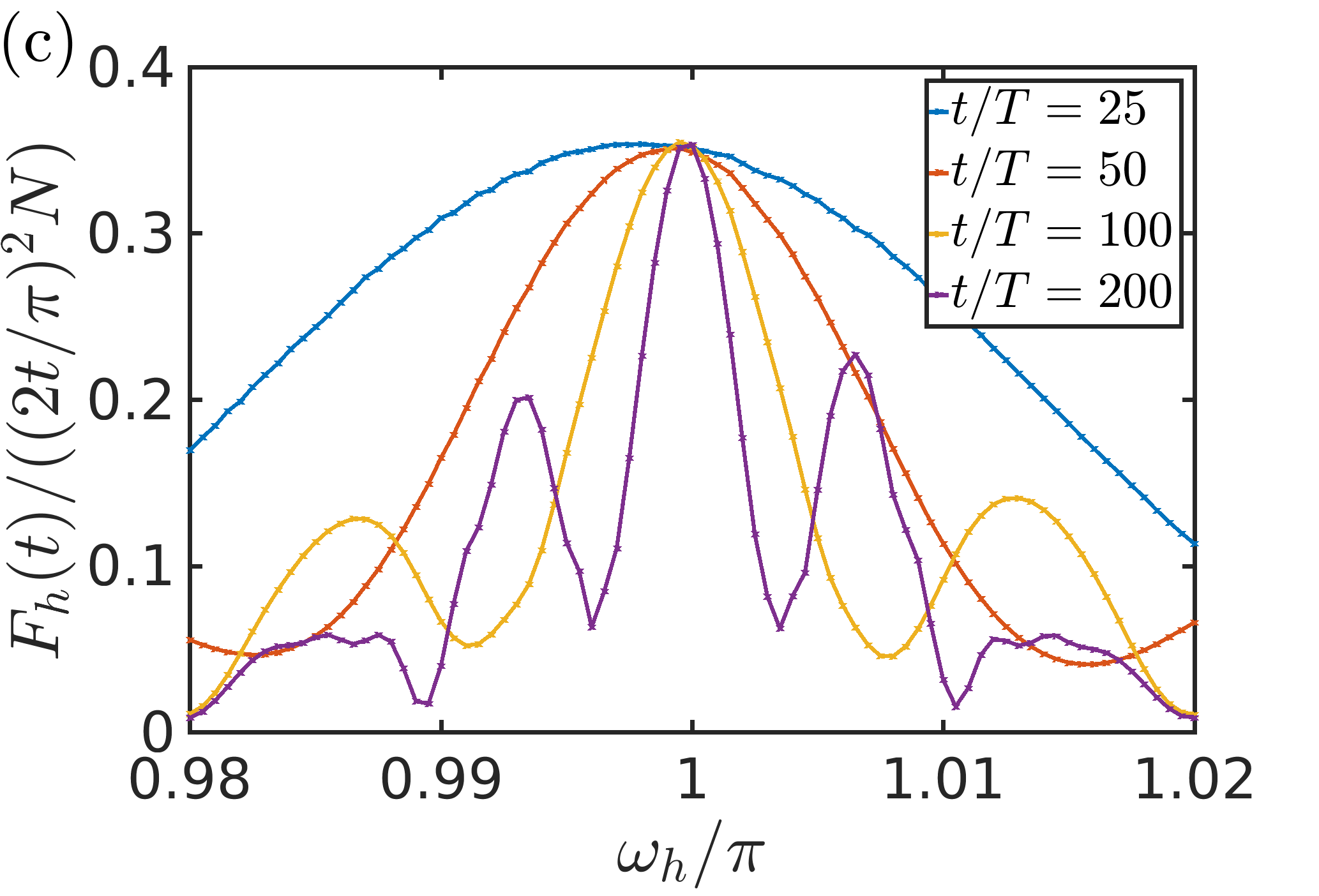}
 \caption{ We show the dynamics of the magnetization and quantum Fisher information for varying AC frequencies.
 We set the sensor parameters on its FTC phase:  $N=5$, $J=b^z =0.25$, $b^x=0.025$, $\theta = 0$, $T = 1$, $\phi=2.8$ and $h =0.1$.
 In (a) we show the magnetization dynamics, where we see that mismatching frequencies tend to destroy the stability of the period doubling dynamics. (b) The effects on the Fisher information is slowing down its growth. In (c) we show the structured dependence of the Fisher information with the AC frequency, for varying fixed times.
 }
 \label{fig.numerical.hacneq0.varwac}
\end{figure*}

 \textbf{Finite $h$, dependence with $\omega_h$: }
 We first show the FTC sensor dependence with the AC frequency, out of the linear response limit. In Fig.\eqref{fig.numerical.hacneq0.varwac} we show such a dependence for a fixed finite $h=10^{-1}$ amplitude field. We see that mismatching AC frequency to the internal FTC one tend to destroy the stability of the period doubling magnetization dynamics - Fig.(\ref{fig.numerical.hacneq0.varwac}a). The time for which these effects occur is inversely proportional to the frequencies mismatching.
 The effects on the Fisher information dynamics - Fig.(\ref{fig.numerical.hacneq0.varwac}b) - is slowing down its growth, till at later times it recovers the quadratic one. Nevertheless, the Fisher information preserves a structured dependence on the AC frequencies for different fixed finite times - Fig.(\ref{fig.numerical.hacneq0.varwac}c).

\begin{figure*}
\includegraphics[scale=0.22]{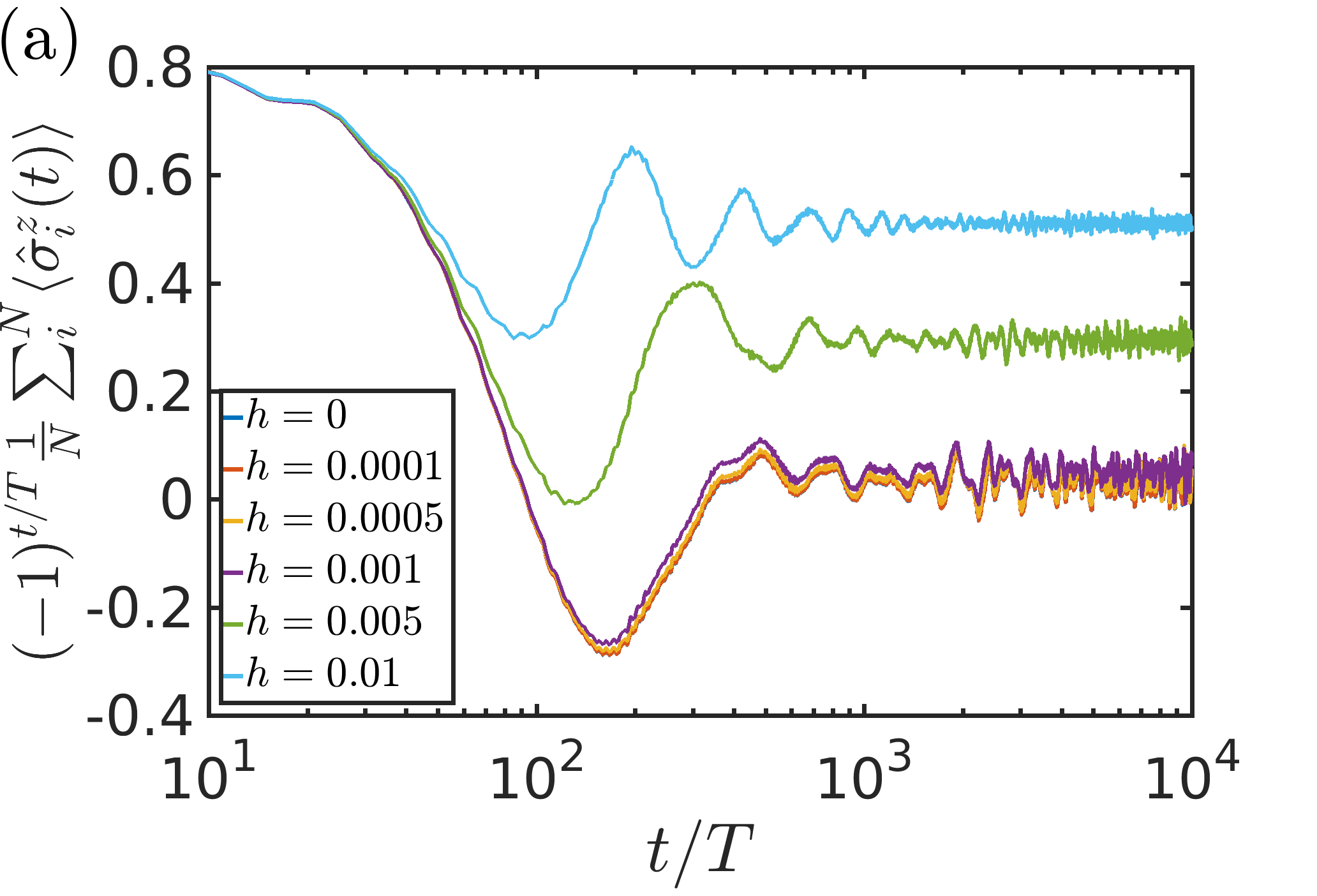}
\includegraphics[scale=0.22]{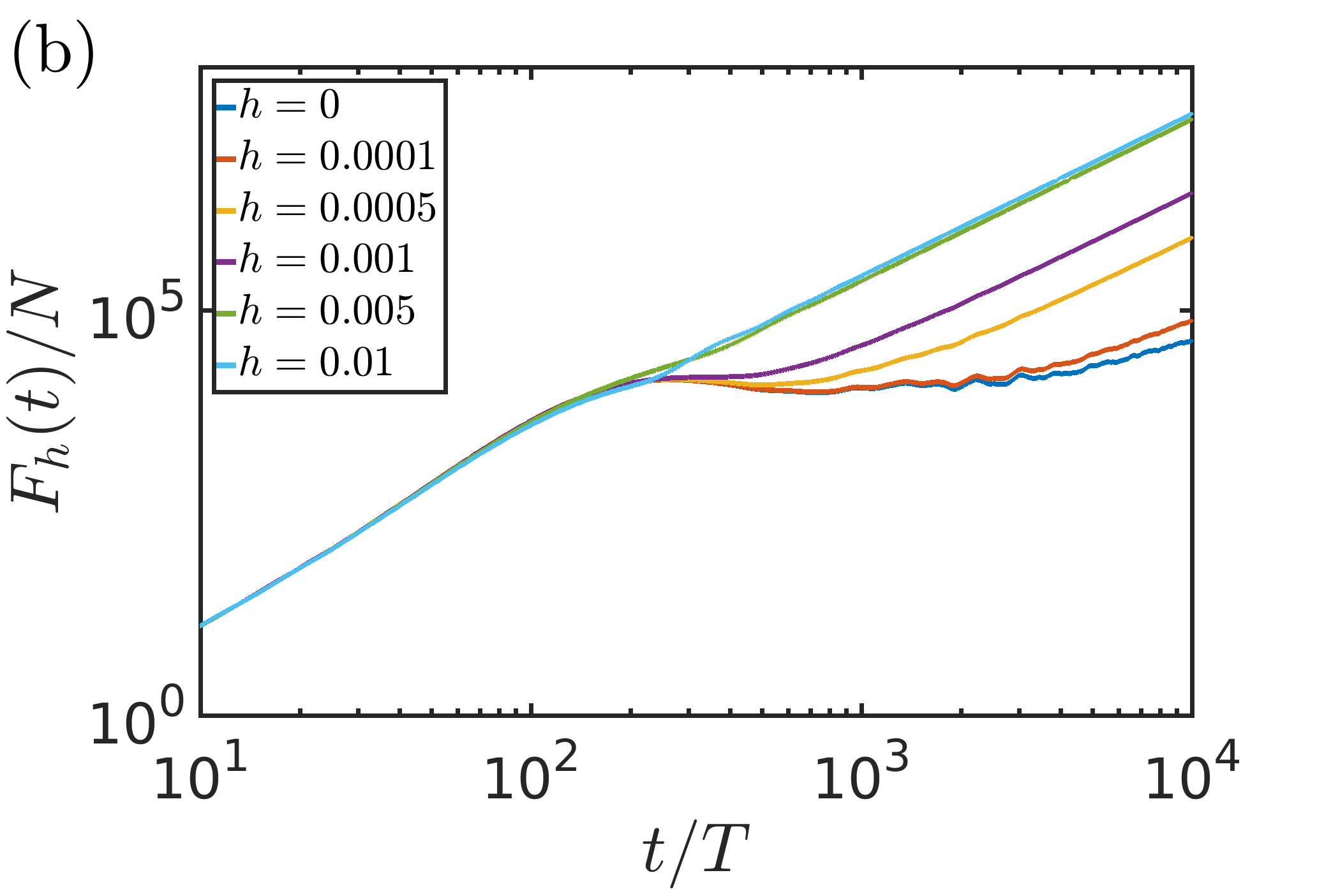}
 \caption{  Dependence with $h$:
 We set a system with $N=4$ spins on its FTC phase, with parameters $J=b^z =0.25$, $b^x=0.025$, $\theta = 0$, $T = 1$, $\phi=2.8$ in the resonant condition $\omega_h =  \pi/T$ and varying AC field $h$.
   In (a)  we show the magnetization dynamics and in (b) the corresponding quantum Fisher information dynamics.
 }
 \label{com.fig.numerical.hacneq0.varhac}
\end{figure*}

\textbf{Dependence with $h$: }
 The sensor dependence with the amplitude field is shown in Fig\eqref{com.fig.numerical.hacneq0.varhac}. We see that dominant AC effects appear in the dynamics at the characteristic time scale $t_{\rm AC} \sim h^{-1}$. For the sensor tuned on period-doubling resonance to the AC field, their main effects tends to stabilize the period-doubling magnetization and recover the quadratic growth in time of the Fisher information.

\section*{Ergodic sensor}

\begin{figure*}
\includegraphics[width = 0.5 \linewidth]{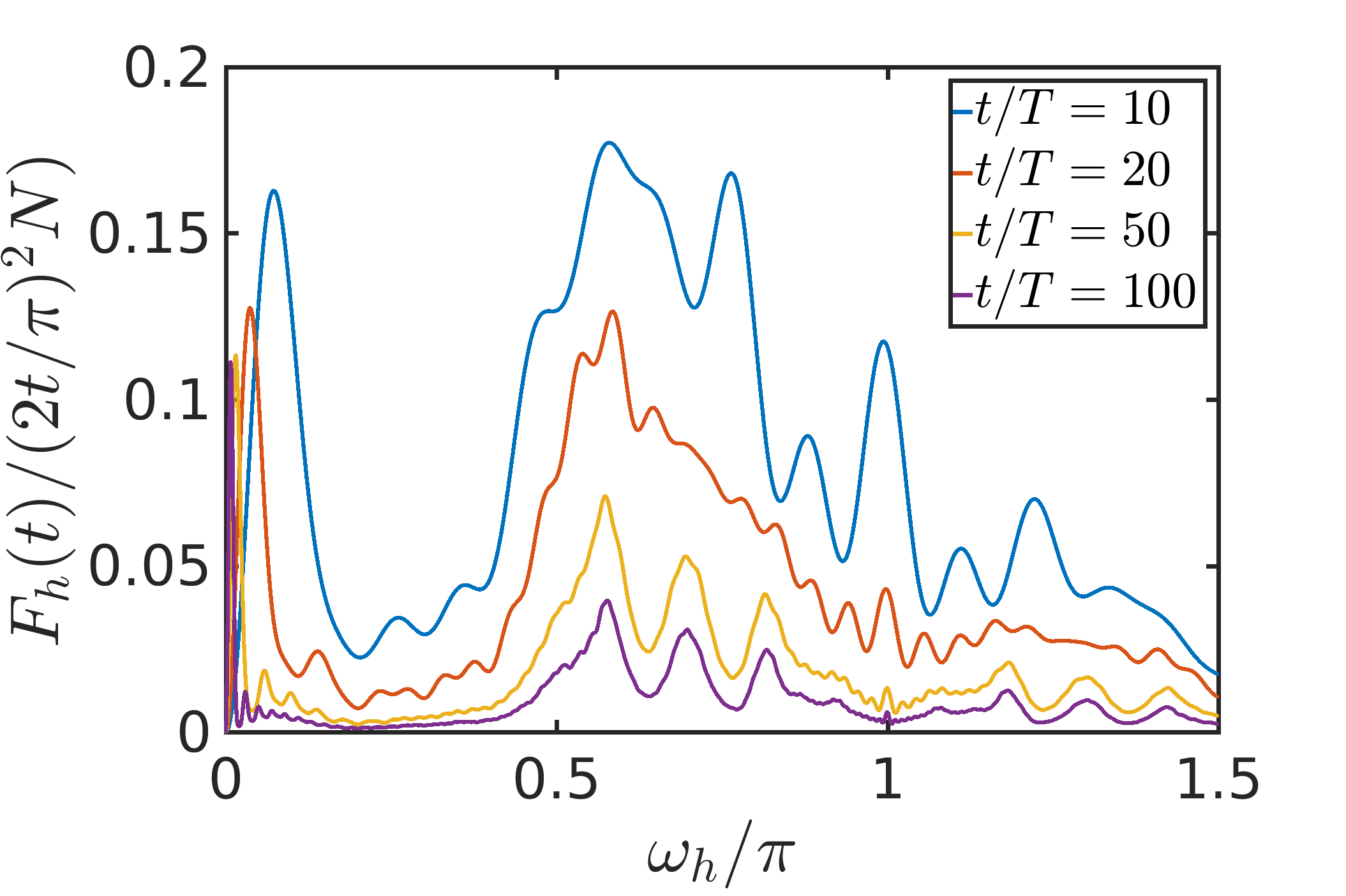}
 \caption{  Ergodic sensor.- We show the QFI response
to the AC-field in the ergodic phase with system parameters $\phi = 2$,
  $J=b^z =0.25$, $b^x=0.025$, $\theta = 0$, $T = 1$ and $h \rightarrow  0$, for a system with $N=4$ spins at different times. The noisy   response is upper bounded by the SQL.
 }
 \label{fig.sensor.ERG}
\end{figure*}

Once tuned out of the FTC phase the sensor loses its long-range spatial time ordering and therefore its improved performance.  We show in Fig\eqref{fig.sensor.ERG} the sensor on its ergodic phase. In this case
the thermalization time is of the order of a few kicks, and roughly independent on the system size. This is reflected in the QFI dynamics, which has no structured response to the probe field, rather a noisy and structureless response, moreover  with a growth in
time slower than quadratic.

\section*{Role of initial states}

In the simulations presented in the main text we used initial separable states $|\psi_{\rm sep} (0)\rangle = [\cos(\vartheta/2) |\uparrow \rangle + e^{i \varphi}\sin(\vartheta/2)|\downarrow\rangle ]^{ \otimes N}$ for the dynamics, with $\vartheta \in [0, 0.05 \pi/2]$ and
$\varphi \in [0, 2\pi]$ random phases.
In this section we expand the discussion on the role of different initial states in the sensing protocol.

While in the main text we mainly focused on uncorrelated initial states (from a practical view the simplest to prepare experimental) we discuss here the possibility of using initial entangled states acting as initial resources for the metrological protocol and their corresponding performance according to the QFI. In the simpler case for the model, with $\phi = \pi$, $b_i^x = 0$,  $h \rightarrow 0$
 we showed analytically in the main text that superlinear enhacements with $N$ can be obtained by preparing entangled initial states, i.e. with superlinear variance along the $z$-direction. In this way we focus here in the general case. Specifically, we consider initial GHZ-like entangled states,
\begin{equation}
\label{eq.ghz.like.states}
|\psi (0,h_{\rm{ac}} \rangle = \left( |\psi_{\rm{sep}} \rangle + g_{\rm{ghz}} \hat X |\psi_{\rm{sep}} \rangle \right) /\sqrt{2}
\end{equation}
where $\hat X = \prod_{i=1}^N \hat \sigma_i^x $ performs a global $\pi$ flip over all spins around the x-direction.
The factor $g_{\rm{ghz}}$ tunes the initial entanglement in the spins, with $g_{\rm{ghz}}=0$ corresponding to the separable states discussed in the main text and $g_{\rm{ghz}}=1$ maximally entangled ones.

\begin{figure*}
\includegraphics[width = 0.45 \linewidth]{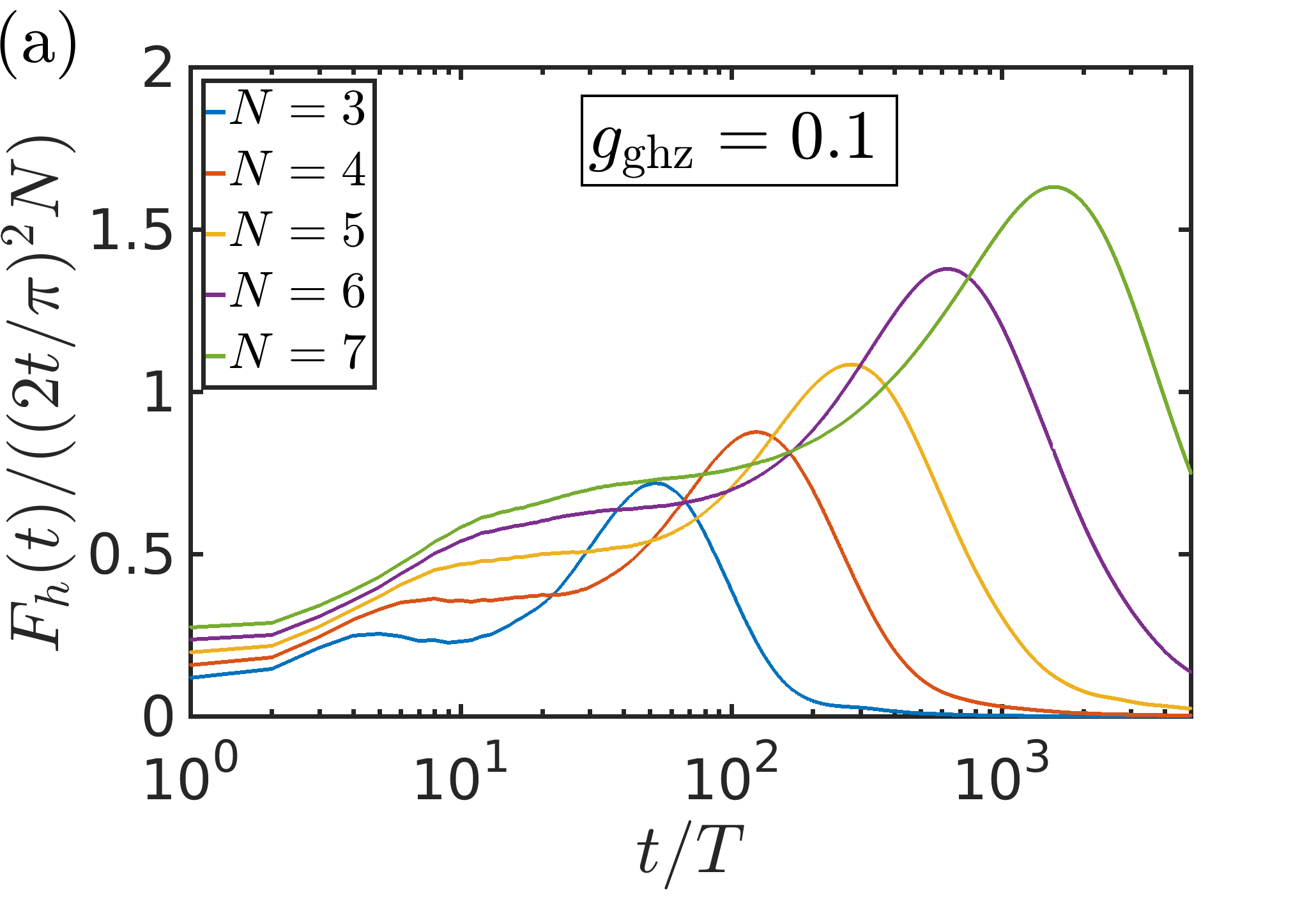}
\includegraphics[width = 0.45 \linewidth]{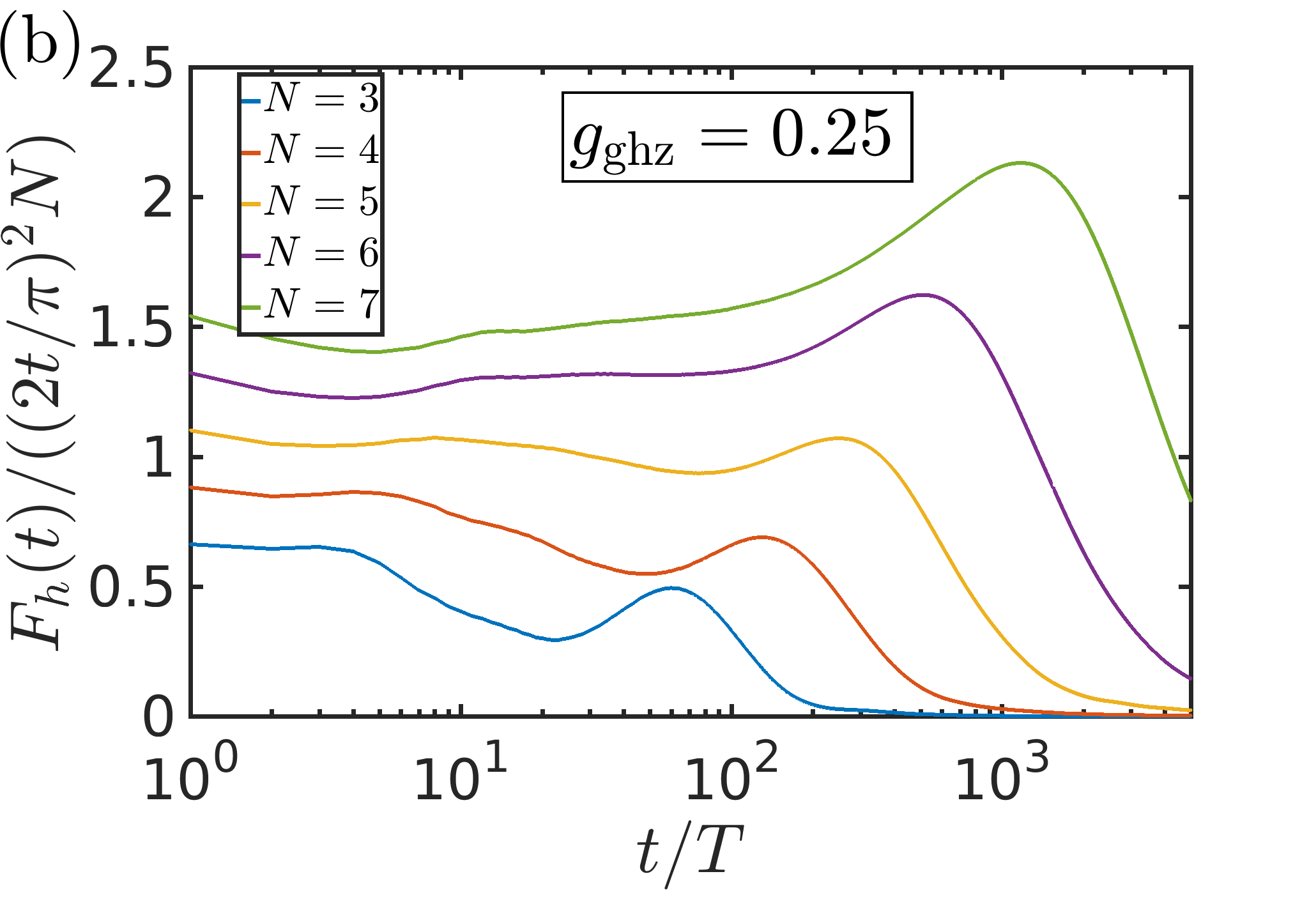}
\includegraphics[width = 0.45 \linewidth]{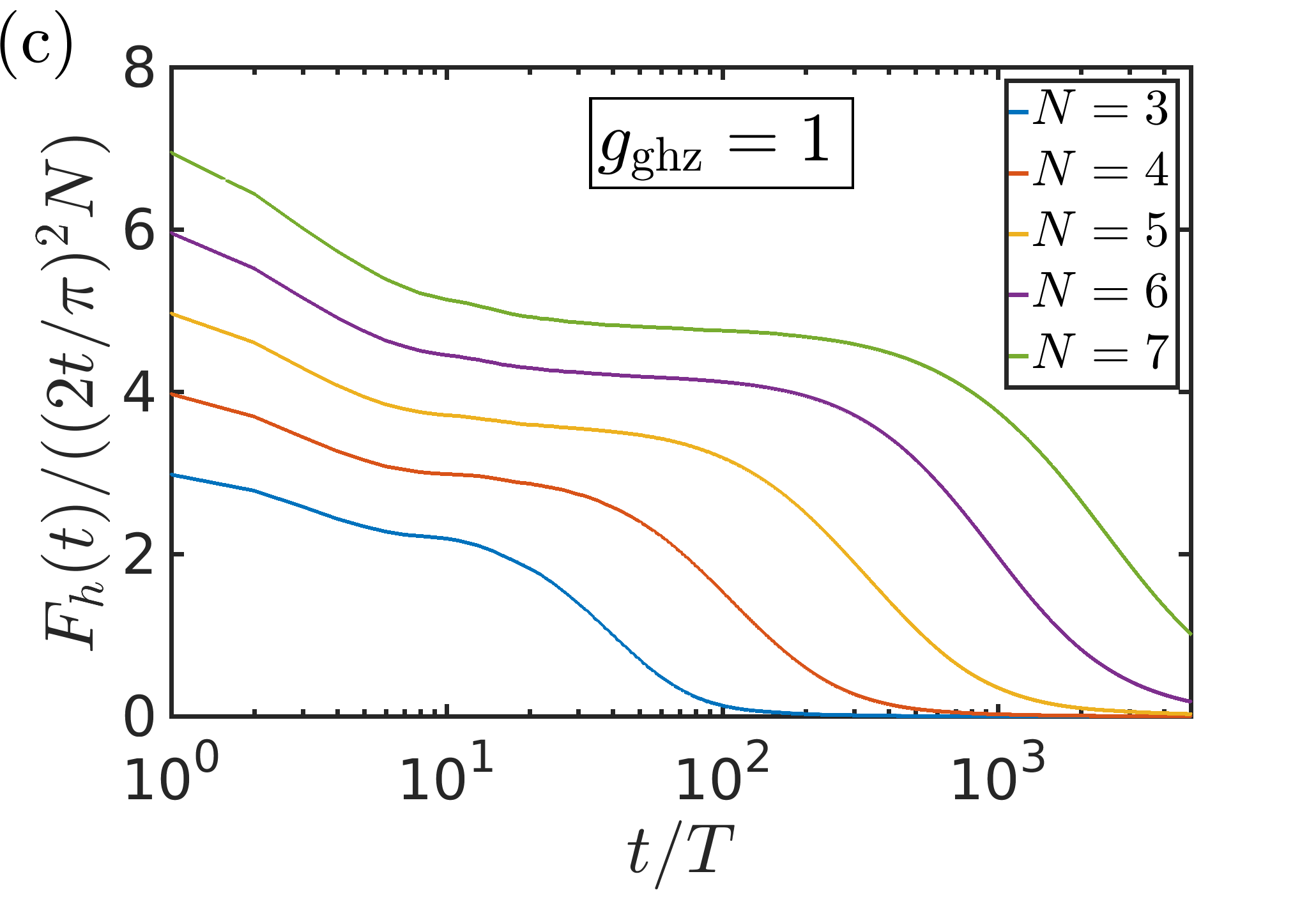}
 \caption{  \textit{Role of initial states.}  Dynamics for the QFI with different GHZ-like states (Eq.\eqref{eq.ghz.like.states}), with the sensor on its FTC phase (same parameters as Fig.($2$) in the main text: $J=b^z =0.25$, $b^x=0.025$, $\theta = 0$, $\phi =2.8$, $T = 1$, $\omega_{h}=\pi/T$ and $h \rightarrow  0$). We show the QFI dynamics with (a) $g_{\rm ghz} = 0.1$,
  (b) $g_{\rm ghz} = 0.25$ and
  (c) $g_{\rm ghz} = 1$, for varying system sizes.
  In all cases the QFI for the initial state is superlinear in $N$. For low-entangled initial states we see usual behavior discussed in the main manuscript, with the emergence of a plateau ($\sim t^2$) for sufficiently large system sizes and a subsequent growth till a peak in the QFI at the thermalization time.
 }
 \label{fig.initial.states}
\end{figure*}

 We show our results in Fig.\eqref{fig.initial.states}. The QFI for low-entangled initial states, with $g_{\rm{ghz}} \lesssim 0.25$ (Figs.(\ref{fig.initial.states}a-b),  shows a qualitative similar behavior as those uncorrelated initial states, i.e., after an initial transient time, they reach a plateau saturating the QSL growth in time ($\sim t^2$), after which displays a growth overcoming it due to correlations spreading. The enhanced performance for the sensor lasts exponentially with the system size, i.e., till the thermalization time. Importantly, the QFI scales superlinearly in $N$ along these different characteristic times, therefore the superlinear enhancements can be preserved and even boosted by the dynamics.
On the other side high-entangled states, usually more fragile to decoherence, show a different behavior with a continuous decreasing in time, neverthless still with different decay rates signaling the different regimes in the period-doubling dynamics. Despite the continuous deteriorization of the sensor in this case, it is interesting noticing such a slow rate which
could still be beneficial for a metrological protocol.

\end{document}